\documentclass[aps,prb,twocolumn,superscriptaddress,10pt,showpacs]{revtex4-1}

\usepackage[utf8x]{inputenc}
\usepackage{graphicx}
\usepackage{amsmath,amsthm,amssymb}
\usepackage[dvipsnames]{xcolor}
\usepackage[colorlinks,linkcolor=blue,citecolor=blue,urlcolor=blue]{hyperref}

\renewcommand{\vec}[1]{\mathbf{#1}}

\newcommand{\hc}{\hat{c}}

\newcommand{\hH}{\hat{H}}

\newcommand{\hrho}{\hat{\rho}}
\newcommand{\brho}{\boldsymbol{\rho}}

\newcommand{\bSig}{\boldsymbol{\Sigma}}
\newcommand{\bG}{{\bf G}}

\begin{document}
\title{Ultrafast nonequilibrium evolution of excitonic modes in semiconductors}
\author{Yuta Murakami}
\affiliation{Department of Physics, Tokyo Institute of Technology, Meguro, Tokyo 152-8551, Japan}
\affiliation{Department of Physics, University of Fribourg,
Fribourg 1700, Switzerland}
\author{Michael Sch\"uler}
\affiliation{Stanford Institute for Materials and Energy Sciences (SIMES),
  SLAC National Accelerator Laboratory, Menlo Park, CA 94025, USA}
\author{Shintaro Takayoshi}
\affiliation{Max Planck Institute for the Physics of Complex Systems,
Dresden 01187, Germany}
\author{Philipp Werner}
\affiliation{Department of Physics, University of Fribourg,
Fribourg 1700, Switzerland}

\date{\today}

\begin{abstract}
We study the time evolution of excitonic states after photo-excitation in the one-dimensional spinless extended Falicov-Kimball model. Several numerical methods are employed and benchmarked against each other: 
time-dependent mean-field simulations, the second-Born approximation (2BA) within the Kadanoff-Baym formalism, the generalized Kadanoff-Baym Ansatz (GKBA) implemented with the 2BA and the infinite time-evolving block decimation (iTEBD) method. 
It is found that the GKBA gives the best agreement with iTEBD and captures the relevant physics.
We find that excitations to the particle-hole continuum and resonant excitations of the equilibrium exciton result in a qualitatively different dynamics.
In the former case, the exciton binding energy remains positive and the frequency of the corresponding coherent oscillations is smaller than the band gap. 
On the other hand, resonant excitations trigger a collective mode whose frequency is larger than the band gap.
We discuss the origin of these different behaviors 
by evaluating the nonequilibrium susceptibility using the nonthermal distribution and a random phase approximation.
The peculiar mode with frequency larger than the band gap is associated with a partial population inversion with a sharp energy cutoff.
We also discuss the effects of the cooling by a phonon bath.
We demonstrate the real-time development of coherence in the polarization, which indicates excitonic condensation out of equilibrium.
\end{abstract}

\maketitle

\section{Introduction}
Excitonic states play a central role in photo-excited semiconductors, nanostructures and molecules and have been studied extensively in the context of photo-voltaic applications~\cite{haug_quantum_1990,ostroverkhova_organic_2016,scholes_excitons_2006,koch_semiconductor_2006} and charge migration.\cite{hill_charge-separation_2000,falke_coherent_2014,bostrom_charge_2018} 
In particular, two-dimensional (2D) materials -- especially transition metal chalcogenides (TMCs) -- are currently attracting a lot of interest, fueled by the possibility of creating tailored heterostructures.\cite{novoselov_two-dimensional_2005,radisavljevic_single-layer_2011,heine_transition_2015,novoselov_2d_2016} 
Due to the low dimensionality of TMCs, the Coulomb interaction is weakly screened, thus giving rise to pronounced interaction effects and excitonic features. TMCs exhibit large exciton binding energies, which can be of the order of a few hundred meV.\cite{he_tightly_2014,heine_transition_2015,cudazzo_exciton_2016} Apart from the importance of excitons as excited states dominating the in-gap optical absorption -- known as virtual or coherent excitons~\cite{haug_quantum_1990,schafer_semiconductor_2013,koch_semiconductor_2006} -- excitons can also be present in the ground state. For sufficiently large binding energy, these excitons can condense collectively, forming an excitonic insulator (EI).\cite{kohn1967,jerome1967,halperin1968} Because of the strong Coulomb interaction, TMCs are among the best candidates for realizing the EI phase.\cite{Monney2007,Hellmann2012NatCom,Wakisaka2009,Kaneko2013,Lu2017}

While virtual excitons in semiconductors are usually considered in the linear response regime, stronger excitations and out-of-equilibrium dynamics have also been in the spot light.
Dynamics of semiconductors after strong excitations and the realization of  the EI phase there have been investigated theoretically,\cite{Comte1986,Ostreich1993,Perfetto2019PRM} and the relevant photo-dressed states have been observed recently.\cite{Murotani2019PRL}
Furthermore, the strong light-matter coupling in TMCs,\cite{britnell_strong_2013} which can be enhanced by orders of magnitude in a micro-cavity setup,\cite{peter_exciton-photon_2005,bisht_universal_2018,latini_cavity_2019} implies that excitonic properties need to be investigated beyond linear response. 
Important examples for nonequilibrium setups also include the optical Stark effect~\cite{sie_valley-selective_2015} and the ultrafast charge transfer in photo-excited bilayer TMCs.\cite{ceballos_ultrafast_2014,Hong2014} 
In addition, in order to unravel the mechanisms of the photo-induced enhancement~\cite{Mor2017PRL,Murakami2017PRL} or ultrafast melting of EI orders~\cite{Okazaki2018NatCom}, it is essential to develop an understanding of the dynamics of bound electron-hole pairs in strongly photo-excited systems.\cite{Hellmann2012NatCom,Denis2016PRB,Mor2017PRL,Murakami2017PRL,Tanabe2018PRB,Tanaka2018PRB,Okazaki2018NatCom} 

In the linear response regime, excitons are typically treated within the framework of the Bethe-Salpeter equation (BSE)~\cite{rohlfing_electron-hole_1998,albrecht_textitab_1998,cudazzo_exciton_2016} in combination with the kernel determined by the Hartree-Fock self-energy (the random-phase approximation, RPA) or the $GW$ approximation.
Extending the BSE to a nonequilibrium scenario is possible,\cite{perfetto_nonequilibrium_2015} but currently out of reach for realistic systems. Time-dependent approaches are a promising alternative route for computing the linear~\cite{berghauser_analytical_2014,perfetto_first-principles_2016,Murakami2016PRBa,Murakami2016PRBb,sangalli_ab-initio_2018} and beyond-linear response.\cite{attaccalite_real-time_2011,attaccalite_two-photon_2018} In particular, the nonequilibrium Green's functions (NEGF)~\cite{stefanucci_nonequilibrium_2013} approach provides a natural way of extending the many-body perturbation theory to the time domain. However, \emph{a priori} it is unclear which scheme works best out of equilibrium. For instance, the spurious effects of fully self-consistent $GW$~\cite{grumet_beyond_2018} are expected to hamper the excitonic properties, while the extension of partially self-consistent schemes to the time domain is not straightforward. Therefore, benchmarks of different methods in and out of equilibrium will yield valuable insights.

In this work, we study a two-band semiconductor model in one dimension (the extended Falicov-Kimball model), with virtual excitons induced by a local inter-band interaction. This simple model has all the ingredients needed for exploring exciton dynamics far from equilibrium, and highly accurate solutions can be obtained. In particular, we employ the infinite time-evolving block decimation (iTEBD)~\cite{Vidal2007PRL} method, which -- upon convergence -- yields an essentially numerically exact description. Furthermore, we employ several methods within the NEGF framework, including time-dependent mean-field (tdMF) theory and the full treatment of the Kadanoff-Baym equations (KBEs).\cite{stefanucci_nonequilibrium_2013} 
The self-energy is treated in the second-Born approximation (2BA), which can capture polarization and exchange effects.
Furthermore, we employ the generalized Kadanoff-Baym ansatz,\cite{lipavsky_generalized_1986} which reduces the computational cost considerably. 
While the iTEBD method is a numerically powerful and reliable method for one-dimensional systems, 
it is difficult to extend the method to more general 
setups such as higher dimensions and long-range interacting systems.  
In the present study, we use it to calculate  
benchmark results for the other methods. Such a systematic comparison for finite systems demonstrated the potential of the GKBA.\cite{schlunzen_nonequilibrium_2017} Here, we will show that the GKBA also performs well in extended systems. 

Benchmarking these methods against each other, we systematically study the properties of excitons out of equilibrium and discuss the effects which require a treatment beyond the MF theory.  
In particular, we compare above-bandgap excitations to resonant excitations of the exciton.
We show that in the latter case a moderately strong pulse can induce large coherent oscillations in the polarization, which can gradually decay and be regarded as a transient nonequilibrium excitonic phase.
We systematically study the nature of collective modes in transient states after above-bandgap excitations and resonant excitations and 
find that in the latter case its nature is different from the normal exciton states in equilibrium.
Combining the GKBA and the RPA-like approach with distributions obtained from GKBA,
we reveal that the peculiar collective mode originates from the efficient creation of an inverted population at the edge of the valence and conduction band.
We also study the cooling effects from the electron-phonon couplings and show the real time formation of the peculiar mode from the above-bandgap excitation and the build-up of an exciton condensation out-of equilibrium.

The paper is organized as follows. In Sec.~\ref{sec:formalism}, we introduce our model and the methods (tdMF, 2BA, GKBA  and iTEBD) used to study the time evolution of the model after photo-excitation. We also derive the expressions for the relevant susceptibilities.
In Sec.~\ref{sec:results}, we show the results of the simulations.
Section~\ref{sec:results_linear} presents the results in the linear response regime,
while in Sec.~\ref{sec:results_nonlinear} we go beyond the linear response regime and discuss the difference between 
above-bandgap excitations and resonant excitations. 
In Sec.~\ref{sec:el_ph_effects}, we consider the effects of cooling from the electron-phonon coupling.
The conclusions of our study are summarized in Sec.~\ref{sec:conclusion}.

\section{Formulation} \label{sec:formalism}
\subsection{Model}
In this paper, we focus on a spinless two-band model,
\begin{align}
\hat{H}(t)&=\hat{H}_{\rm kin}(t)+\hat{H}_{\rm int}+\hat{H}_{\rm dip}(t) \ ,
\end{align}
where the first term represents the kinetic energy
\begin{align}
\hat{H}_{\rm kin}&=-\sum_{\langle i,j\rangle ,a=c,v} J_a({\bf r}_{ij},t) \hc^\dagger_{i,a} \hc_{j,a}+\sum_{i,a} \Delta_a \hc^\dagger_{i,a}\hc_{i,a} \ .
\end{align}
Here $\langle i,j\rangle$ indicates a pair of nearest-neighbor sites, and $a=c,v$ indicates the orbitals. 
$c$ and $v$ stand for the conduction band and the valence band, respectively.
$\hc^\dagger$ is the electron creation operator, $J_a({\bf r}_{ij},t)$ the hopping parameter, ${\bf r}_{ij}$ is the spatial vector connecting site $j$ to site $i$, and $\Delta_a$ is the energy of orbital $a$.
The electrons in the two bands interact via a local interaction 
\begin{align}
\label{eq:hint}
\hH_{\rm int}&=U\sum_i \hat{n}_{i,c}\hat{n}_{i,v},
\end{align}
where $\hat{n}_{i,a} = \hc_{i,a}^\dagger \hc_{i,a}$.
The effect of an external field is partially included in $J_a({\bf r}_{ij},t)$ via the Peierls substitution 
\begin{align}
J_a({\bf r}_{ij},t)=J_a({\bf r}_{ij})\exp\Big[i\frac{q}{\hbar} {\bf r}_{ij} \cdot {\bf A}(t)\Big],
\end{align}
where ${\bf A}(t)=- \int^t {\bf E}(\bar{t}) d\bar{t}$ is the vector potential, ${\bf E}(t)$ is the electric field,
and $q$ the charge of the electron. This term corresponds to the intraband acceleration.
The third term is the dipole excitation, which represents the interband excitation,
\begin{align}
\hat{H}_{\rm dip}(t)=-{\bf E}(t) \cdot \sum_{i}\hat{\bf P}_i=-{\bf E}(t)\cdot \sum_{i,a} {\bf d}_{a} \hc^\dagger_{i,a} \hc_{i,\bar{a}}.
\end{align}
Here the dipole matrix $\vec{d}_{c,v}$  is local and $\hat{\bf P}$ is the dipole moment per site. 
We use the notation $\bar{a} = c$ ($\bar{a} = v$) for $a=v$ ($a=c$).
In the following, we set the length of the primitive vector, $\hbar$ and $q$ to unity.

Assuming translational invariance, we define the operators in momentum space, $\hc^\dagger_{\vec{k},a} = \frac{1}{\sqrt{N}} \sum_{\bf k} e^{i {\bf k}\cdot {\bf r}_i} \hc^\dagger_{i,a}$. 
Here $N$ is the number of sites.
With these operators, one can express the Hamiltonian as 
\begin{subequations}
\begin{align}
\hH_{\rm kin}(t) &= \sum_{\bf k} 
\begin{bmatrix}
\hc^\dagger_{{\bf k} ,c} & \hc^\dagger_{{\bf k} ,v} 
\end{bmatrix}
\cdot 
{\bf h}_{{\rm kin},{\bf k} }(t)
\cdot 
\begin{bmatrix}
\hc_{{\bf k} ,c} \\ \hc_{{\bf k} ,v} 
\end{bmatrix},
\\
\hH_{\rm dip}(t) &= \sum_{\bf k} 
\begin{bmatrix}
\hc^\dagger_{{\bf k} ,c} & \hc^\dagger_{{\bf k} ,v} 
\end{bmatrix}
\cdot 
{\bf h}_{{\rm dip},{\bf k} }(t)
\cdot 
\begin{bmatrix}
\hc_{{\bf k} ,c} \\ \hc_{{\bf k} ,v} 
\end{bmatrix},
\end{align}
\end{subequations}
with 
\begin{subequations}
\begin{align}
{\bf h}_{{\rm kin},{\bf k} }(t) & = 
\begin{bmatrix}
\epsilon_c({\bf k}-q{\bf A}(t)) +\Delta_c  & 0 \\
0 &  \epsilon_v({\bf k}-q{\bf A}(t)) +\Delta_v
\end{bmatrix},\\
{\bf h}_{{\rm dip},{\bf k} }(t)& =
\begin{bmatrix}
0 & -{\bf E}(t)\cdot {\bf d}_c    \\
-{\bf E}(t)\cdot {\bf d}_v  & 0 
\end{bmatrix}.
\end{align}
\end{subequations}
Here $\epsilon_a({\bf k}) = -\sum_l J_a({\bf r}_{l})e^{-i{\bf k}\cdot {\bf r}_l}$, 
where the sum runs over nearest-neighbor sites.

Next we introduce the single-particle density matrix as 
\begin{subequations}\label{eq:rho_k}
\begin{align}
\rho_{ia,jb}(t)&\equiv \langle \hc^\dagger_{jb}(t)\hc_{ia}(t)\rangle\\
\rho_{{\bf k}, a,b}(t)&\equiv \langle \hc^\dagger_{{\bf k},b}(t)\hc_{{\bf k},a}(t)\rangle.
\end{align}
\end{subequations}
Note that $\rho_{{\rm loc},a,b} (t) \equiv \rho_{ia,ib}(t) = \frac{1}{N} \sum_{\bf k}  \rho_{{\bf k}, a,b}(t)$.
We also use $\boldsymbol{\rho}_{\bf k}(t)$ to express the $2\times 2$ matrix with elements $\rho_{{\bf k}, a,b}(t)$.

In the present study, we consider one-dimensional chains and assume that the dipole matrix is directed along the chain and that ${\bf d}_c^* = {\bf d}_v$.
The system is excited with Gaussian pulses with various excitation frequencies.

\subsection{Methods}
In order to study the nonequilibrium dynamics of this system, we use several different methods: tdMF, the 2BA, the GKBA implemented with the 2BA and the iTEBD. In the following, we briefly introduce these methods and discuss the corresponding susceptibilities. 

In general, a linear function $\chi^R_{\rm BA}(t,t')= -i\theta(t-t') \langle [\hat{B}(t),\hat{A}(t')]\rangle$ can be measured by exciting the system with a weak excitation, $\hat{H}_{\rm ex}=F_{\rm ex}(t)\hat{A}$ 
with $F_{\rm ex}(t) \propto \delta(t-t')$,
and observing the evolution of $\hat{B}$. This is how we measure linear functions in the following.
If $\hat{A}=\sum_{ij}A_{ij} \hat{c}^\dagger_j\hat{c}_i$  and $\hat{B} = \sum_{ml} B_{ml} \hat{c}^\dagger_l \hat{c}_m$, the response function can be expressed as
\begin{align}
\chi^R_{BA} (t,t') = \sum_{ijlm} B_{ml} \chi^R_{ml,ij}(t,t') A_{ij},
\end{align}
where $\chi^R_{ml,ij}(t,t')$ is the retarded part 
of the function 
\begin{align} \label{eq:chi_mlij}
\chi_{ml,ij}(t,t') =& -i \langle {\mathcal T}_{\mathcal C} \hat{c}^\dagger_l(t) \hat{c}_m(t)\hat{c}^\dagger_j(t')\hat{c}_i (t')\rangle \\
& + i \langle {\mathcal T}_{\mathcal C} \hat{c}^\dagger_l(t) \hat{c}_m(t)\rangle \langle {\mathcal T}_{\mathcal C} \hat{c}^\dagger_j(t')\hat{c}_i (t')\rangle\nonumber
\end{align}
defined on the
the Konstantinov-Perel' contour $\mathcal{C}$,\cite{Konstantinov1961,stefanucci_nonequilibrium_2013} which runs from time $0$ to time $t_\text{max}$ along the real time axis, back to zero, and then to $-i\beta$ along the imaginary time (Matsubara) axis. 
${\mathcal T}_{\mathcal C}$ is the contour ordering operator and $t,t^\prime \in \mathcal{C}$ refer to contour arguments. 

In particular, we consider the response function for 
 $\hat{A} =  \rho_{\nu,j} \equiv \hat{\Psi}^\dagger_j \boldsymbol{\sigma}_\nu \hat{\Psi}_j $ and $\hat{B} = \rho_{\mu,i} \equiv \hat{\Psi}^\dagger_i \boldsymbol{\sigma}_\mu \hat{\Psi}_i$, which we denote by
  $\chi^R_{\mu\nu}(t-t';{\bf r}_{ij})$ for a steady state. 
 Here $\hat{\Psi}_i = [\hc_{i,c}\;\;\hc_{i,v}]^T$ and $\boldsymbol{\sigma}_\mu$ is a Pauli matrix.
In momentum space this response function is expressed as 
$\chi^R_{\mu\nu}(\omega;{\bf q})=\sum_l  \int dt e^{i\omega t}\chi^R_{\mu\nu}(t;{\bf r}_{l})e^{-i{\bf q}\cdot {\bf r}_{l}}$.
Here, $\chi^R_{11}$ corresponds to the polarization-polarization response function.

\subsubsection{Time-dependent mean-field theory}
In the tdMF theory, we consider the time evolution of the one-particle density matrix Eq.~\eqref{eq:rho_k} under the MF Hamiltonian, which is self-consistently determined at each time.
Assuming translational invariance, the MF Hamiltonian is
\begin{align} \label{eq:H_MF1}
\hH_{\rm MF}(t) = \sum_{\bf k} 
\begin{bmatrix}
\hc^\dagger_{{\bf k},c} & \hc^\dagger_{{\bf k},v} 
\end{bmatrix}
\cdot 
{\bf h}_{{\rm MF},{\bf k}}(t)
\cdot 
\begin{bmatrix}
\hc_{{\bf k},c} \\ \hc_{{\bf k},v} 
\end{bmatrix},
\end{align}
with 
\begin{subequations}\label{eq:H_MF2}
\begin{eqnarray}
{\bf h}_{{\rm MF},{\bf k}}(t) &=& {\bf h}_{{\rm kin},{\bf k}}(t) + {\bf h}_{{\rm Hartree},{\bf k}}(t)\nonumber\\
&& + {\bf h}_{{\rm Fock},{\bf k}} (t) 
+ {\bf h}_{{\rm dip},{\bf k}}(t), \\
{\bf h}_{{\rm Hartree},{\bf k}}(t) &=& U
\begin{bmatrix}
 \rho_{{\rm loc},vv}(t)  & 0 \\
0 &  \rho_{{\rm loc},cc}(t) 
\end{bmatrix}, \label{eq:Hartree}\\
{\bf h}_{{\rm Fock},{\bf k}} &=& -U
\begin{bmatrix}
0 & \rho_{{\rm loc},cv}(t)   \\
\rho_{{\rm loc},vc}(t)  & 0 
\end{bmatrix}.\label{eq:Fock}
\end{eqnarray}
\end{subequations}
The time evolution of the density matrix follows from the van Neumann equation, $\partial_t \boldsymbol{\rho}_{\bf k}(t) = -i[{\bf h}_{{\rm MF},{\bf k}}(t),\boldsymbol{\rho}_{\bf k}(t) ]$ and the MF effect is taken into account through $\boldsymbol{\rho}_{{\rm loc}}(t)= \frac{1}{N} \sum_{\bf k} \boldsymbol{\rho}_{{\bf k}}(t)$. 
We also note that the Hartree term shifts the positions of the bands after the excitation since the occupation in the two orbitals changes.

Now we consider the linear response of a \emph{steady solution} in the MF dynamics assuming that the steady state does not break the symmetry of the Hamiltonian (the system remains in the normal state).
Here a steady solution means a state which does not change under the MF time propagation.
The equilibrium state is one example.
The expression for $\chi^R_{\mu\nu}(\omega;{\bf q})$ evaluated by the direct time propagation within the tdMF 
is 
\begin{align}
\boldsymbol{\chi}^R(\omega;{\bf q}) = [1-\boldsymbol{\chi}^R_0(\omega;{\bf q})\boldsymbol{\Theta}]^{-1}\boldsymbol{\chi}^R_0(\omega;{\bf q}).\label{eq:rpa_like}
\end{align}
Here $\boldsymbol{\chi}$ indicates the $2\times2 $ matrix whose components are $\chi^R_{\mu\nu}$ with $\mu,\nu=1,2$,
and $\boldsymbol{\Theta}={\rm diag}[-\frac{U}{2},-\frac{U}{2}]$.
$\boldsymbol{\chi}_0$ is the response evaluated by the time evolution without updating the mean field, 
which can be expressed as 
\begin{align}
\chi_{0,\mu\nu}(t;{\bf q}) = -i\theta(t) \frac{1}{N} &\sum_k 
\bigl\{ {\rm tr} [\boldsymbol{\sigma}_\mu \boldsymbol{\mathcal{G}}^>_{\bf k+q}(t) \boldsymbol{\sigma}_\nu \boldsymbol{\mathcal{G}}^<_{\bf k}(-t)] \nonumber \nonumber\\
& - {\rm tr} [\boldsymbol{\sigma}_\mu \boldsymbol{\mathcal{G}}^<_{\bf k+q}(t) \boldsymbol{\sigma}_\nu \boldsymbol{\mathcal{G}}^>_{\bf k}(-t)]\bigl\}.\label{eq:chi0_noneq}
\end{align}
Here $\boldsymbol{\mathcal{G}}_{\vec{k}}(t)$ is the MF Green's function at the steady-state, which is expressed as 
\begin{subequations}\label{eq:g_noneq}
\begin{align}
\mathcal{G}^<_{aa,{\bf k}}(t) &= i n_{a}({\bf k}) e^{-iE_{a}({\bf k}) t}, \\
\mathcal{G}^>_{aa,{\bf k}}(t) &= -i (1-n_{a}({\bf k})) e^{-iE_{a}({\bf k}) t},
\end{align}
\end{subequations}
with vanishing off-diagonal components, since we assume that the steady state is a normal state. 
$E_{a}({\bf k})$ is the energy of the electron in band $a$ with momentum ${\bf k}$ 
determined with the MF Hamiltonian, Eq.~\eqref{eq:H_MF2}, for the density distribution $n_a({\bf k})$.
The explicit expression of the Fourier transformation of $\boldsymbol{\chi}_0(t;{\bf q})$ is 
\begin{align}\label{eq:chi0_noneq_w}
\chi_{0,\mu\nu}(\omega;{\bf q}) = \frac{1}{N} 
\sum_{{\bf k},a,b}
\frac{ {\rm tr}[\boldsymbol{W}_a \boldsymbol{\sigma}_\mu \boldsymbol{W}_b \boldsymbol{\sigma}_\nu ] (n_a({\bf k-q})-n_b({\bf k}))}{\omega + i0^+ -(E_{b}({\bf k})-E_{a}({\bf k-q}))},
\end{align}
with $\boldsymbol{W}_c = \begin{bmatrix} 1 & 0 \\ 0 & 0 \end{bmatrix}$ and $\boldsymbol{W}_v = \begin{bmatrix} 0 & 0 \\ 0 & 1 \end{bmatrix}$.
We note that by using the equilibrium distribution $n_a({\bf k})=(1+\exp(\beta E_a({\bf k})))^{-1}$, 
Eq.~\eqref{eq:rpa_like} reproduces the well-known RPA-type susceptibility in equilibrium, which consists of ladder diagrams, see Appendix \ref{sec:Tmat}.

One can simplify Eq.~(\ref{eq:chi0_noneq_w}) for ${\bf q}={\bf 0}$ by introducing 
\begin{align}
\boldsymbol{\gamma} = \boldsymbol{L}^{-1} \boldsymbol{\chi} \boldsymbol{L}, \;\;  \boldsymbol{\gamma}_0 = \boldsymbol{L}^{-1} \boldsymbol{\chi}_0 \boldsymbol{L}, \;\; 
\boldsymbol{L}=\frac{1}{\sqrt{2}}
\begin{bmatrix}
1 & i \\
i  & 1
\end{bmatrix}.
\end{align}
This rotation makes the off-diagonal elements of $\boldsymbol{\gamma}$ and $\boldsymbol{\gamma}_0$ zero, while  
\begin{subequations}\label{eq:gamma_noneq_w}
\begin{align}
\gamma_{0,11}(\omega) &= 
\frac{2}{N} 
\sum_{\bf k}
\frac{n_c(k)-n_{v}({\bf k})}{\omega + i0^+ -(E_{v}({\bf k})-E_{c}({\bf k}))},\\
\gamma_{0,22}(\omega) &= 
\frac{2}{N} 
\sum_{{\bf k}}
\frac{n_v({\bf k})-n_c({\bf k})}{\omega + i0^+ -(E_c({\bf k})-E_v({\bf k}))},\label{eq:gamma022_noneq_w}\\
\gamma_{\mu\mu} (\omega)&= 
\frac{\gamma_{0,\mu\mu}(\omega)}{1+\frac{U}{2}\gamma_{0,\mu\mu}(\omega)}. \label{eq:gamma_rpa}
\end{align}
\end{subequations}
We note that for positive frequencies ($\omega>0$), $\gamma_{11} (\omega)$ and $\gamma_{0,11} (\omega)$ are featureless, while 
$\gamma_{22} (\omega)$ and $\gamma_{0,22} (\omega)$ are responsible for nontrivial features in $\boldsymbol{\chi}$ and $\boldsymbol{\chi}_0$.
In particular, $\chi_{11}=\frac{1}{2}(\gamma_{22} + \gamma_{11} )$ implies that $\chi_{11}$ and  $\gamma_{22}$ exhibit similar structures.

\subsubsection{Full Kadanoff-Baym formalism: Second-Born approximation}

In order to investigate the out-of-equilibrium correlated dynamics beyond the tdMF approximation, higher-order scattering processes need to be taken into account. The NEGF framework~provides a systematic and versatile approach for treating many-body effects in the time domain. \cite{stefanucci_nonequilibrium_2013,Aoki2013} We define the general Green's function GF on the Konstantinov-Perel' contour $\mathcal{C}$ as
\begin{align}
\label{eq:def_gf}
G_{ab,\vec{k}}(t,t^\prime) = -i \langle \mathcal{T}_{\mathcal{C}} \hat{c}_{\vec{k},a}(t)
\hat{c}^\dagger_{\vec{k},b}(t^\prime) \rangle  .
\end{align} 
Adopting again the matrix notation, the GF obeys the equation of motion (Dyson equation)
\begin{align}
	\label{eq:dyson}
	\left[i \partial_t -{\bf h}_{\mathrm{MF},\vec{k}}(t)\right] {\bf G}_{\vec{k}}(t,t^\prime) &= 
	\delta_\mathcal{C}(t,t^\prime) + [\boldsymbol{\Sigma}_{\mathrm{corr},\vec{k}} \ast {\bf G}_{\vec{k}}](t,t^\prime) \ ,
\end{align}
where $\delta_\mathcal{C}(t,t^\prime)$ is a straightforward generalization of the Dirac delta function to the contour $\mathcal{C}$, while $\ast$ denotes the convolution along $\mathcal{C}$. Solving Eq.~\eqref{eq:dyson} is accomplished by projecting onto observable times by invoking the Langreth rules, yielding the KBEs.\cite{stefanucci_nonequilibrium_2013,Aoki2013} After solving the corresponding equilibrium state (Matsubara GF), the real-time evolution is governed by the KBEs. 
Since the MF self-energy $\bSig_{\rm HF}(t,t')=\delta_\mathcal{C}(t,t')({\bf h}_{\mathrm{Hartree}}(t)+ {\bf h}_{\mathrm{Fock}}(t))$ is included in ${\bf h}_{\mathrm{MF}}(t)$, many-body effects beyond mean field are captured by the correlation self-energy $\bSig_{\mathrm{corr}} = \bSig_{\mathrm{corr}}[\bG]$, which is a functional of the GF. In this work, we employ the 2BA, which corresponds to the second-order self-consistent weak-coupling approximation: 
$\bSig_{\mathrm{corr}}[\bG]\approx \bSig^{\mathrm{2B}}[\bG]$.
The correlated parts of the self-energy consists of a direct and and an exchange part,
\begin{align}
\label{eq:sigma_2b}
\bSig^{\mathrm{2B}}[\bG](t,t^\prime) = \bSig^{\mathrm{2Bd}}[\bG](t,t^\prime) + \bSig^{\mathrm{2Bx}}[\bG](t,t^\prime) \ .
\end{align}
For the interaction Hamiltonian~\eqref{eq:hint}, the direct contribution to the self-energy reads
\begin{align}
\label{eq:sigma2b_d}
\Sigma^{\mathrm{2Bd}}_{ab,\vec{k}}(t,t^\prime) &= \frac{U^2}{N^2}\sum_{\vec{q},\vec{p}} 
G_{ab,\vec{k}-\vec{q}}(t,t^\prime) G_{\bar{a}\bar{b},\vec{q}+\vec{p}}(t,t^\prime) \nonumber \\ &\quad \times G_{\bar{b}\bar{a},\vec{p}}(t^\prime,t) \ ,
\end{align}
while the exchange part is given by
\begin{align}
\label{eq:sigma2b_x}
\Sigma^{\mathrm{2Bx}}_{ab,\vec{k}}(t,t^\prime) &= -\frac{U^2}{N^2}\sum_{\vec{q},\vec{p}} 
G_{a\bar{b},\vec{k}-\vec{q}}(t,t^\prime) G_{\bar{a}b,\vec{q}+\vec{p}}(t,t^\prime) \nonumber\\ &\quad \times  G_{\bar{b}\bar{a},\vec{p}}(t^\prime,t) \ .
\end{align}
While exchange effects captured by Eq.~\eqref{eq:sigma2b_x} vanish when the GFs do not have inter-orbital components, 
their impact onto the strongly driven dynamics is less clear. 
Therefore, we also compare results within the simplified 2BA (taking the direct contribution Eq.~\eqref{eq:sigma2b_d}) to the full 2BA.
We denote the simplified 2BA 
as s2BA in the following.

Given the expression of the self-energy, one can evaluate the linear response functions by simulating the evolution after a weak delta-function field pulse. 
Using the real-space representation for convenience, 
the corresponding response function ($\chi_{ml,ij}(t,t')$ in Eq.~\eqref{eq:chi_mlij}) can be expressed as 
\begin{align}
\label{eq:chi_contour}
&\chi_{ml,ij}(t,t')  = -i  {\rm tr}[{\bf e}_{lm} \bG_0(t,t') {\bf e}_{ji} \bG_0(t',t)]
\nonumber\\
& -i {\rm Tr}\big[{\bf e}_{lm}  \int_C d\bar{t}_1 d\bar{t}_2 \bG_0(t,\bar{t}_1)  \frac{\delta_\mathcal{C} \bSig[\bG](\bar{t}_1,\bar{t}_2) }{\delta_\mathcal{C} F_{\rm ex}(t';i,j)}\Bigl|_{{\bf G}={\bf G}_0} \bG_0(\bar{t}_2,t)\big].
\end{align}
Here, $\bG_0$ indicates the full Green's function without the probe excitation, $F_{\rm ex}(t;i,j)$ is the strength of the external field proportional to $\hc^\dagger_j \hc_i$, 
$\frac{\delta_\mathcal{C}}{\delta_\mathcal{C}}$ is the functional derivative on the contour, and $\frac{\delta_\mathcal{C} \bSig[\bG](\bar{t}_1,\bar{t}_2) }{\delta_\mathcal{C} F_{\rm ex}(t';i,j)}$ the reducible vertex expressed as a functional derivative on the contour $\mathcal{C}$. The matrix ${\bf e}_{ij}$ is defined by $[{\bf e}_{ij}]_{kl} = \delta_{ik}\delta_{jl}$. The self-energy $\bSig[\bG]$ entering Eq.~\eqref{eq:chi_contour} is the full self-energy $\bSig[\bG] = \bSig_{\mathrm{HF}}[\bG] + \bSig_{\mathrm{corr}}[\bG]$.
We note that the contribution from $\frac{\delta_\mathcal{C} \bSig^{F}[\bG](\bar{t}_1,\bar{t}_2) }{\delta_\mathcal{C} F_{\rm ex}(t';i,j)}$ leads to the ladder diagrams consisting of $G_0$.
In other words, the response to the probe evaluated by only updating $\bSig^{F}[\bG]$ in the Dyson equation and keeping $\bSig_{\rm corr}[\bG]= \bSig_{\rm corr}[\bG_0]$ corresponds to the ladder diagrams consisting of $G_0$.
Hence, $\frac{\delta_\mathcal{C} \bSig_{\rm corr}[\bG](\bar{t}_1,\bar{t}_2) }{\delta_\mathcal{C} F_{\rm ex}(t';i,j)}$ 
generates diagrams beyond these ladder diagrams.

\subsubsection{Generalized Kadanoff-Baym ansatz} 
The numerical cost of evaluating the full Kadanoff-Baym equations Eq.~\eqref{eq:dyson} scales as $\mathcal{O}(N\cdot N_t^3)$, where $N_t$ is the number of time points used in the simulation, and it grows significantly for long propagation times. Employing the GKBA reduces the computational effort by one order of magnitude in $N_t$ and thus allows simulations up to considerably longer times. Furthermore, the GKBA has been shown to cure some deficiencies of the full KBE approach, especially for finite systems.\cite{schlunzen_nonequilibrium_2017} Systematic assessments in extended system are scarce,\cite{schuler_quench_2018} which is one of the motivations for the present study.

Within the GKBA, the description is reduced to the time evolution of the single-particle density matrix. 
Given a self-consistent approximation to the self-energy ($\bSig=\bSig[\bG]$), the equation of motion for the density matrix (transport equation) can be expressed as
\begin{align}
\label{eq:KBE_transport}
\partial_t \brho_{\vec{k}}(t) + i [{\bf h}_{\mathrm{MF},\vec{k}}[\brho](t),\brho_{\vec{k}}(t)] = - ({\bf I}^<_{\vec{k}}(t,t)+\mathrm{h.\,c.}), 
\end{align}
where the collision integral ${\bf I}^<_{\vec{k}}(t,t)$ is defined by
\begin{align}
\label{eq:collint}
{\bf I}^<_{\vec{k}}(t,t) &= \int^t_{-\infty}\! d\bar{t} \big( \bSig^<_{\mathrm{corr},\vec{k}}(t,\bar{t}) \bG^A_{\vec{k}}(\bar{t},t) \nonumber \\ &\quad + \bSig^R_{\mathrm{corr},\vec{k}}(t,\bar{t}) 
\bG^<(\bar{t},t) \big) .
\end{align}
Here, we consider the Keldysh contour, which starts from $t=-\infty$, in constrast to the Konstantinov-Perel' contour used in the previous section. 
In the Keldysh formalism,  
correlations of the initial state $\brho(t=0)$ are built in by adiabatic switching: at $t=-\infty$, the equilibrium density matrix is determined by the MF treatment, while correlation effects are gradually incorporated by replacing $\bSig_{\mathrm{corr},\vec{k}}(t,t^\prime)\rightarrow f(t)f(t^\prime)\bSig_{\mathrm{corr},\vec{k}}(t,t^\prime)$ with a smooth switch-on function $f(t)$.
However, Eqs.~\eqref{eq:KBE_transport} and \eqref{eq:collint} are not closed in terms of $\brho$ since, in principle, information on the whole two-time dependence of the GF enters the collision integral Eq.~\eqref{eq:collint}.

The idea of the GKBA is to approximate the Green's functions (GF) in the collision integral by 
combining the information contained in the occupation
(${\brho}$)
and the 
spectrum ($\tilde{\bG}^R,\tilde{\bG}^A$) by introducing the following auxiliary GF:
\begin{subequations}
\begin{align}
\tilde{\bG}^<_{\vec{k}}(t,t') &= - \tilde{\bG}^R_{\vec{k}}(t,t') \brho_{\vec{k}}(t') + \brho_{\vec{k}}(t) {\tilde{\bG}}^A_{\vec{k}}(t,t'), \\
\tilde{\bG}^>_{\vec{k}}(t,t') &= \tilde{\bG}^R_{\vec{k}}(t,t') (1- \brho_{\vec{k}}(t')) - (1- \brho_{\vec{k}}(t)) {\tilde{\bG}}^A_{\vec{k}}(t,t').
\end{align}
\end{subequations}
Here we determine $\tilde{\bG}^{R/A}(t,t')$ as the mean-field GF
\begin{align}
(i\partial_t - {\bf h}_{\rm HF}[\brho](t))  \tilde{\bG}^{R/A}(t,t') = \delta(t-t') \ .
\end{align}
The GKBA attains a closed form for any choice of the self-energy upon replacing $\bSig[\bG] \rightarrow \bSig[\tilde{\bG}]$ and $\bG \rightarrow \tilde{\bG}$ in the collision integral Eq.~\eqref{eq:collint}.
In the present paper, we use the full 2BA Eq.~\eqref{eq:sigma_2b} as well as the simplified version which considers the direct contribution Eq.~\eqref{eq:sigma2b_d} only (s2BA). 

We now roughly discuss the relation between the susceptibility evaluated by GKBA and the full KBE form as described in the previous section.
As mentioned in the previous section, keeping $\bSig_{\rm corr}[\bG]= \bSig_{\rm corr}[\bG_0]$ in the full KBE corresponds to the ladder diagram in terms of the full GF $\bG_0$, which is in contrast 
to the tdMF, whose ladder diagram consists of the MF GF. In the latter GF, the damping of quasi-particles is not included. 
In the language of the transport equation, Eq.~\eqref{eq:KBE_transport}, this corresponds to keeping  $\bSig_{\rm corr}[\bG]= \bSig_{\rm corr}[\bG_0]$
but updating ${\bG}$ in the collision integral.
In the GKBA we approximately update $\bG$ and $\bSig$ in the collision integral.
Therefore, naively speaking, the corresponding susceptibility should include a) the effects of the ladder diagrams consisting of dressed Green's function (more than the MF Green's function) and b) the effects beyond the ladder diagrams.

\subsubsection{iTEBD}

In this subsection, we briefly explain the principle of 
iTEBD.\cite{Vidal2007PRL}  
This method can be applied for 
the time-dependent problems 
such as quench dynamics or laser driving
in quantum spin\cite{Barmettler2009PRL,Takayoshi2014PRB,Takayoshi2019PRB}
and fermion~\cite{Bauer2015PRB, Ono2016PRB,Coulthard2017PRB,Ono2017PRB}
systems. 
The advantage of iTEBD is that calculations 
without finite size effects are possible 
by assuming translational invariance of the system.

In one dimension, the quantum states can be represented as 
matrix product states (MPS). 
When the system has a translational symmetry, 
the MPS representation is also translationally invariant
\begin{align}
 |\Psi\rangle=&\sum_{\alpha_{i},s_{i}}\cdots
   \lambda_{\alpha_{-1}}^{BA}
   \Gamma_{\alpha_{-1}\alpha_{0}}^{A}[s_{0}]
   \lambda_{\alpha_{ 0}}^{AB}
   \Gamma_{\alpha_{ 0}\alpha_{1}}^{B}[s_{1}]
   \lambda_{\alpha_{ 1}}^{BA}\nonumber\\
   &\times\Gamma_{\alpha_{ 1}\alpha_{2}}^{A}[s_{2}]
   \lambda_{\alpha_{ 2}}^{AB}\cdots
   |\ldots,s_{0},s_{1},s_{2},\ldots\rangle,
\nonumber
\end{align}
where $s_{i}$ represents the quantum state on the site $i$, 
and in the present system $s_{i}=0,1,2,3$ correspond to 
$(n_{iv},n_{ic})=(0,0),(1,0),(0,1),(1,1)$, respectively 
($n_{iv},n_{ic}$ is the eigenvalue of $\hat{n}_{iv},\hat{n}_{ic}$). 
$\alpha_{i}$ is the suffix for the matrices, and 
the values in the diagonal matrix $\lambda_{\alpha_{i}}$ 
($=\lambda_{\alpha_{i}\alpha_{i}}$) are singular values 
(also known as the entanglement spectrum) 
obtained from the Schmidt decomposition on the bond 
between the sites $i$ and $i+1$. 
The bipartition of the sites into $A$ and $B$ is 
for the purpose of the time evolution described below. 

The initial state is $s_{i}=1$ for all $i$, 
and the MPS representation is given as 
$\lambda_{\alpha_{i}=1}^{AB(BA)}=1$ and 
$\Gamma_{\alpha_{i}=1,\alpha_{i+1}=1}^{A(B)}[s_{i}]=\delta_{s_{i}1}$, 
where the matrix dimension is 1. 
Next we write the Hamiltonian in the bipartite form as 
\begin{align}
 \hat{H}(t)=\sum_{i\in A}\hat{H}_{i}^{AB}(t)
   +\sum_{i\in B}\hat{H}_{i}^{BA}(t),
\nonumber
\end{align}
where
\begin{align}
 \hat{H}_{i}^{AB(BA)}(t)=&
   -\sum_{a}[J_{a}({\bf r}_{i,i+1},t)\hat{c}_{i,a}^{\dagger}\hat{c}_{i+1,a}
     +\mathrm{H.c}]\nonumber\\
   &+\sum_{a}\frac{\Delta_{a}}{2}
     (\hat{c}_{i,a}^{\dagger}\hat{c}_{i,a}
     +\hat{c}_{i+1,a}^{\dagger}\hat{c}_{i+1,a})\nonumber\\
   &+\frac{U}{2}
     (\hat{n}_{i,c}\hat{n}_{i,v}+\hat{n}_{i+1,c}\hat{n}_{i+1,v})\nonumber\\
   &-\frac{1}{2}\mathbf{E}(t)\cdot
     (\hat{\mathbf{P}}_{i}+\hat{\mathbf{P}}_{i+1}).\nonumber
\end{align}
Note that $\hat{H}^{AB(BA)}(t)$ only acts on the bond $AB(BA)$. 
Using the Trotter formula, the time evolution operator 
$\mathcal{U}(t,t+\Delta t)$
for an infinitesimal time interval from $t$ to $t+\Delta t$
is decomposed as 
\begin{align}
 \hat{U}(t,t+\Delta t)
   =&e^{-i\sum_{i\in A}\hat{H}_{i}^{AB}(t+\frac{\Delta t}{2})\frac{\Delta t}{2}}
     e^{-i\sum_{i\in B}\hat{H}_{i}^{BA}(t+\frac{\Delta t}{2})\Delta t}\nonumber\\
    &\times e^{-i\sum_{i\in A}\hat{H}_{i}^{AB}(t+\frac{\Delta t}{2})\frac{\Delta t}{2}}
   +\mathcal{O}(\Delta t^{2})\nonumber\\
   =&\prod_{i\in A}e^{-i\hat{H}_{i}^{AB}(t+\frac{\Delta t}{2})\frac{\Delta t}{2}}
     \prod_{i\in B}e^{-i\hat{H}_{i}^{BA}(t+\frac{\Delta t}{2})\Delta t}\nonumber\\
    &\times \prod_{i\in A}e^{-i\hat{H}_{i}^{AB}(t+\frac{\Delta t}{2})\frac{\Delta t}{2}}
   +\mathcal{O}(\Delta t^{2}) \ .\nonumber
\end{align}
We can consider 
$\mathcal{T}_{s_{i}s_{i+1};s_{i}'s_{i+1}'}
\equiv e^{-i\hat{H}_{i}^{AB}(t+\frac{\Delta t}{2})\frac{\Delta t}{2}}$ 
as a two-site quantum gate, 
and the procedure of its application is as follows. 
We construct a large matrix 
\begin{align}
 \Theta_{\alpha_{i-1}s_{i}s_{i+1}\alpha_{i+1}}^{AB}
   =&\sum_{\alpha_{i},s_{i}',s_{i+1}'}
   \lambda_{\alpha_{i-1}}^{BA}
   \Gamma_{\alpha_{i-1}\alpha_{i}}^{A}[s_{i}']
   \lambda_{\alpha_{i}}^{AB}\nonumber\\
   \times&\Gamma_{\alpha_{i}\alpha_{1}}^{B}[s_{i+1}']
   \lambda_{\alpha_{i+1}}^{BA}
   \mathcal{T}_{s_{i}s_{i+1};s_{i}'s_{i+1}'},\nonumber
\end{align}
and then perform the singular value decomposition, 
\begin{align}
 \Theta_{\alpha_{i-1}s_{i}s_{i+1}\alpha_{i+1}}^{AB}
   =\sum_{\alpha_{i}'}
   X_{\alpha_{i-1}s_{i}\alpha_{i}'}^{A}
   \tilde{\lambda}_{\alpha_{i}'}^{AB}
   Y_{\alpha_{i}'s_{i+1}\alpha_{i+1}}^{B}
\nonumber
\end{align}
by regarding $(\alpha_{i-1},s_{i})$ and $(s_{i+1},\alpha_{i+1})$ 
as the row and column of the matrix, respectively. 
The number of updated singular values 
$\tilde{\lambda}_{\alpha_{i}'}^{AB}$ is four times larger than 
that of $\lambda_{\alpha_{i}}^{AB}$ 
because $\tilde{\lambda}$ is obtained from the enlarged matrix 
$\Theta_{(\alpha_{i-1},s_{i});(s_{i+1},\alpha_{i+1})}$ 
($s_{i}=0,1,2,3$). 
Since the dimension of the matrix increases by iterating the step, 
we fix a maximum dimension $M$ 
(called the truncation dimension) and only keep the $M$ largest 
singular values, truncating the rest 
when the matrix dimension exceeds $M$.  
The updated $\Gamma$ is constructed as 
\begin{align}
 \tilde{\Gamma}_{\alpha_{i-1}\alpha_{i}'}^{A}[s_{i}]
   =&(\lambda_{\alpha_{i-1}}^{BA})^{-1}
     X_{\alpha_{i-1}s_{i}\alpha_{i}'}^{A},\nonumber\\
 \tilde{\Gamma}_{\alpha_{i}'\alpha_{i+1}}^{B}[s_{i+1}]
   =&Y_{\alpha_{i}'s_{i+1}\alpha_{i+1}}^{B}
   (\lambda_{\alpha_{i+1}}^{BA})^{-1}.\nonumber
\end{align}
The procedure is the same for the application of 
$\prod_{i\in B}e^{-i\hat{H}_{i}^{BA}(t+\frac{\Delta t}{2})\Delta t}$. 
By iterating the above update, 
we can calculate the time evolution of the system. 
The numerical error arises from the Trotter decomposition 
and the truncation, and the precision becomes better 
for larger $M$ and smaller $\Delta t$.
In this paper, we set $M=1200$ and 
$\Delta t=0.01$ or 0.05 depending on the laser field. 
The expectation value of a single-site observable 
such as $\hat{n}_{i,v}$ and $\hat{P}_{i}$ 
(for the $A$ site)
is calculated as 
\begin{align}
 \langle \hat{O}_{i}\rangle
   =&\sum_{\alpha_{i-1},s_{i},\alpha_{i}}
   (\lambda_{\alpha_{i-1}}^{BA})^{2}
   \Gamma_{\alpha_{i-1}\alpha_{i}}^{A*}[s_{i}]
   \Gamma_{\alpha_{i-1}\alpha_{i}}^{A}[s_{i}']
   (\lambda_{\alpha_{i}}^{AB})^{2}\nonumber\\
   &\qquad\times\langle s_{i}|\hat{O}_{i}|s_{i}'\rangle,
\nonumber
\end{align}
where $*$ represents the complex conjugate.
We also calculate the expectation value for the $B$ site 
in the same way and take the average of $A$ and $B$. 

For the calculations of space-time correlation functions, 
we use TEBD for finite size systems instead of iTEBD 
because the application of the single site operator at the initial time $t_{0}$
breaks the spatially translational invariance.
We prepare the $N$ ($=\mathrm{even}$) site system $r=-N/2+1,\ldots,N/2$,
and apply the operator at the site $r=0$. 
The scheme for the time evolution of TEBD is the same as that of iTEBD. 
Hence the response functions are obtained directly 
\begin{align}
\label{eq:chi_iTEBD}
 \underline{\chi}^>(\omega;{\bf q},t_0) &=\int_{t_0}^{t_{1}}dt\sum_{\bf r}  e^{i\omega ((t-t_{0})-{\bf q}\cdot{\bf r})}\chi_{11}^> (t,t_{0};{\bf r} ), 
\end{align}
where 
$\chi_{11}^{>} (t,t_{0};{\bf r}) =-i \langle \hat{P}_{\bf r}(t)\hat{P}_{\bf 0}(t_{0})\rangle$
is the greater part of the contour function $\chi_{11} (t,t';{\bf r}) \equiv -i\langle \mathcal{T}_{\mathcal {C}}\hat{P}_{\bf r}(t)\hat{P}_{\bf 0}(t')\rangle$.
This quantity Eq.~\eqref{eq:chi_iTEBD} reveals the structure of collective modes at finite momenta. 
$\chi_{11}^{>} (t,t_{0};{\bf r})$ can be calculated as follows. 
Since the initial state is $(n_{i,v},n_{i,c})=(1,0)$ for all $i$,
the initial MPS is represented by one-dimensional matrix as stated above.
For the equilibrium correlation function, 
we apply $\hat{P}_{\bf 0}$ to this state ($t_{0}=0$),
and calculate the time evolution using the Hamiltonian without laser
up to the time $t$. Then $\hat{P}_{\bf r}$ is applied 
and taking the inner product with the initial state
(and the phase factor $e^{iE_{0}t}$, $E_{0}$ is the ground state energy).
For the dynamical correlation function under the laser,
we evolute from the initial MPS up to the time $t_{0}$ 
with the Hamiltonian with laser driving and obtain the state $|\Psi(t_{0})\rangle$.
Then we evolute the two states $|\Psi(t_{0})\rangle$ and $\hat{P}_{\bf 0}|\Psi(t_{0})\rangle$
from $t_{0}$ to $t_{1}$ with the Hamiltonian under laser and 
apply $\hat{P}_{\bf r}$ only to the latter.
$\chi_{11}^{>} (t,t_{0};{\bf r})$ is obtained as the inner product
of these two states.

We note that $-\mathrm{Im}\underline{\chi}^<(\omega;{\bf q})$ in equilibrium at $T=0$ is exactly the same as $-\mathrm{Im} \chi^R(\omega;{\bf q})$ for  $\omega>0$. 
In general, when the contribution from the lesser part of $\chi_{11} (t,t';{\bf r})$ is small,  $\underline{\chi}^>(\omega;{\bf q},t_0)$ can be approximated with  the Fourier component of the retarded part $\chi^R(\omega;{\bf q},t_0)$.
In practice we use a window function $F_{\rm window}(t;t_0)$ in the Fourier transformation Eq.~\eqref{eq:chi_iTEBD}, 
$\chi_{11}^>\rightarrow \chi_{11}^> F_{\rm window}$, 
since TEBD can only access rather short times. Specifically, we use $F_{\rm window}(t;t_0)=F_{\rm gauss} (t-t_0,\sigma)$ with 
$F_{\rm gauss} (t,\sigma) \equiv \exp\bigl(-\frac{t^2}{2\sigma^2}\bigl)$.

\section{Results}\label{sec:results}

\begin{figure}
  \centering
    \hspace{-0.cm}
    \vspace{0.0cm}
   \includegraphics[width=60mm]{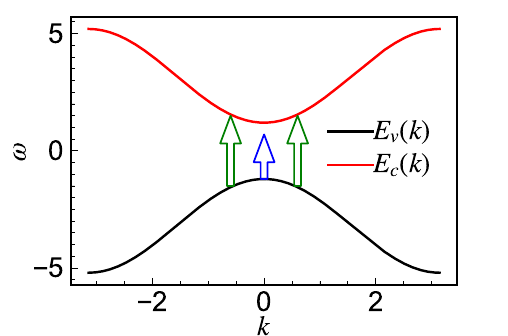} 
  \caption{Dispersion of the conduction band and valence band for $J_c=1,J_v=-1, \Delta_v = -3.2, \Delta_c=1.2$ and $U=2.0$ at $T=0$. 
  The green (blue) arrows indicate the above-gap (resonant) excitation with frequency $\Omega=3.0$ ($\Omega=1.9$).}
  
  \label{fig:dispersion_eq}
\end{figure}

In the following, we choose the hopping parameters as $J_c=1,J_v=-1$ and consider half-filling systems in the semiconductor regime (with a band gap $>0$).
In this case the valence band is fully occupied in the ground state at $T=0$, which is our initial state.  
The single particle spectrum obtained by the MF theory becomes exact for this state, as discussed in Appendix.~\ref{sec:Tmat}.
For the other parameters, we use $\Delta_v = -3.2$ and $\Delta_c=1.2$, and $U=2.0$, which corresponds  
to a direct gap semiconductor with the band gap ${\mathcal E}_{\rm gap}=2.4$ at $T=0$, see Fig.~\ref{fig:dispersion_eq}.
The choice of these parameters is motivated by those of some TMDs, which are characterize by a binding energy of a few hundred meV and a gap energy of a few eV.\cite{Hong2014} 
 
\subsection{Linear response regime} \label{sec:results_linear}
\begin{figure}
  \centering
    \hspace{-0.cm}
    \vspace{0.0cm}
   \includegraphics[width=85mm]{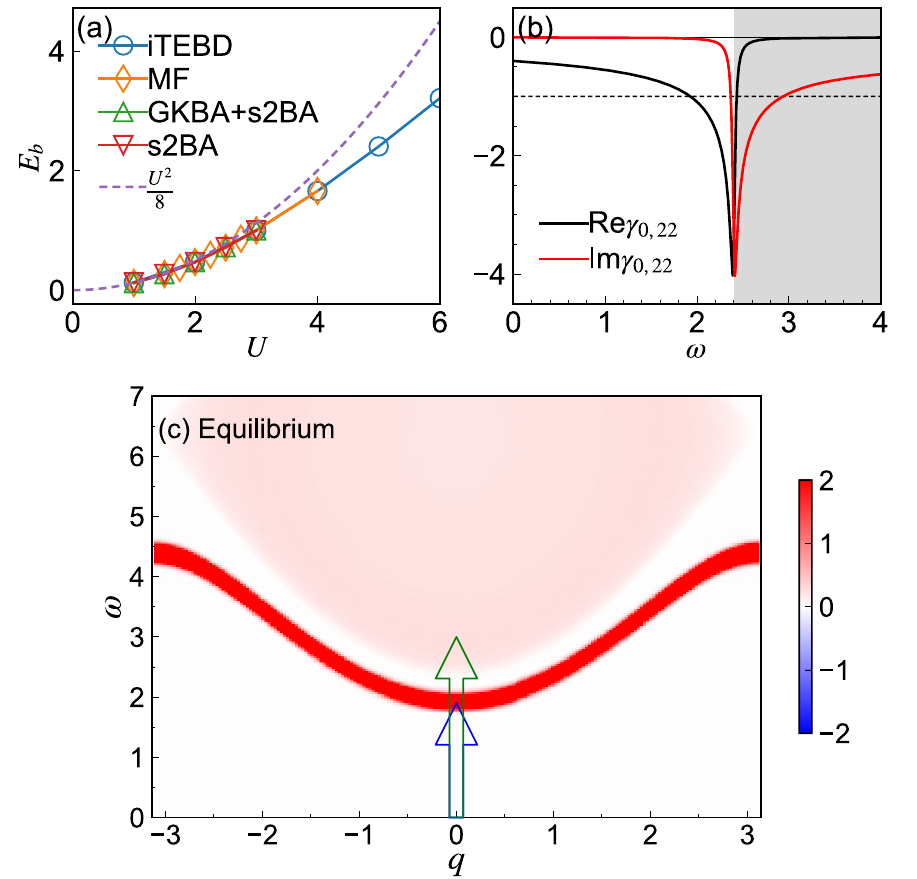} 
  \caption{(a) Comparison of the exciton binding energy, $\mathcal{E}_b$, estimated from the oscillations after a short pulse using different numerical methods. 
  The dashed line indicates $\frac{U^2}{8}$.
  (b) Results of $\gamma_{0,22}$ (Eq.~\eqref{eq:gamma022_noneq_w}) for $T=0$. Here $0^+ = 0.02$ is used. The horizontal dotted line indicates $-\frac{2}{U}$. 
  The shaded area indicates the particle hole continum.
  (c) The spectrum of the linear response function $-{\rm Im}\chi_{11}(\omega;q)$ in equilibrium evaluated by the TEBD for $\Delta_v = -3.2, \Delta_c=1.2$ and $U=2.0$ at $T=0$.
   Here we take $\sigma=(t_1-t_0)/(2\sqrt{2})$ with $t_0=0$ and $t_1=40$.
   The green (blue) arrows indicate the above-gap (resonant) excitation with excitation frequency $\Omega=3.0$ ($\Omega=1.9$). 
  }
  \label{fig:VEffect_HF_Free_comparison}
\end{figure}

We first discuss the excitons in the equilibrium system.  
The exciton state is a bound state of an electron in the conduction band and a hole in the valence band.
When we denote the energy necessary to excite an exciton from the equilibrium state by ${\mathcal E}_{\rm ex}$, the exciton binding energy ${\mathcal E}_{b}$ can be expressed as ${\mathcal E}_{b} ={\mathcal E}_{\rm gap} - {\mathcal E}_{\rm ex}$.
To measure ${\mathcal E}_{\rm ex}$, we excite the system with a very weak and short pulse, which includes a wide range of frequency components, and measure the 
induced dynamics of the dipole moment $P$. 
The exciton energy ${\mathcal E}_{\rm ex}$ manifests itself as a well defined oscillation in this quantity, and thus can be obtained by the Fourier transformation of $P(t)$. 
In Fig.~\ref{fig:VEffect_HF_Free_comparison}(a), we compare the ${\mathcal E}_{b}$ evaluated in the above way for different methods (s2BA, GKBA+s2BA,tdMF, iTEBD).
The results match perfectly, since in the present case one can show that the MF dynamics (RPA-type response), the GKBA and 2B are exact, see Appendix A.
(2BA and GKBA+2BA are also exact.) 
More specifically, the response function evaluated by Eq.~\eqref{eq:gamma_noneq_w} with the $T=0$ occupation becomes exact.
In Fig.~\ref{fig:VEffect_HF_Free_comparison}(b), we show the corresponding $\gamma_{0,22}(\omega)$.
The imaginary part of $\gamma_{0,22} (\omega)$ is essentially zero below the band gap.
(It is finite in the figure because we use $0^+=0.02$ in  Eq.~\eqref{eq:gamma022_noneq_w} for the numerical evaluation.)
The real part has a peak at ${\mathcal E}_{\rm gap}$, which is related to the imaginary part by the Kramers-Kronig relation.
The crossing of $\gamma_{0,22} (\omega)$ and $-2/U$ at $\omega<{\mathcal E}_{\rm gap}$ leads to a peak structure in the imaginary part of $\gamma_{22} (\omega)$, which corresponds to the exciton.
For the one dimensional case, one can analytically show that $-{\rm Re}\gamma_{22,0}(\omega)$ diverges $\propto \frac{1}{\sqrt{\omega-{\mathcal E}_{\rm gap}}}$ around ${\mathcal E}_{\rm gap}$ for $\omega<{\mathcal E}_{\rm gap}$ and that the binding energy ${\mathcal E}_{b}$ scales as $\frac{U^2}{4*(J_c-J_v)}$ for small $U$. 
We also note that, as long as the ground state is semimetallic, the exciton binding energy is independent of ${\mathcal E}_{\rm gap}$ in the present case.
One can see this from Eq.~\eqref{eq:gamma022_noneq_w}. The change of the gap by $\Delta {\mathcal E}_{\rm gap}$ just shifts $\gamma_{0,22}(\omega)$ by $\Delta {\mathcal E}_{\rm gap}$. 
Hence the pole position in $\gamma_{22}(\omega)$ is also shifted by $\Delta {\mathcal E}_{\rm gap}$, and the binding energy does not change.

In Fig.~\ref{fig:VEffect_HF_Free_comparison}(c), we show the spectrum of the linear response function $-{\rm Im}\chi_{11}(\omega;q)$ 
for $\Delta_v = -3.2, \Delta_c=1.2$ and $U=2.0$ at $T=0$ evaluated by the TEBD.
One can see a dispersive band below the particle-hole continuum, which corresponds to the (virtual) exciton states and their dispersion.

 \begin{figure*}
  \centering
    \hspace{-0.cm}
    \vspace{0.0cm}
   \includegraphics[width=170mm]{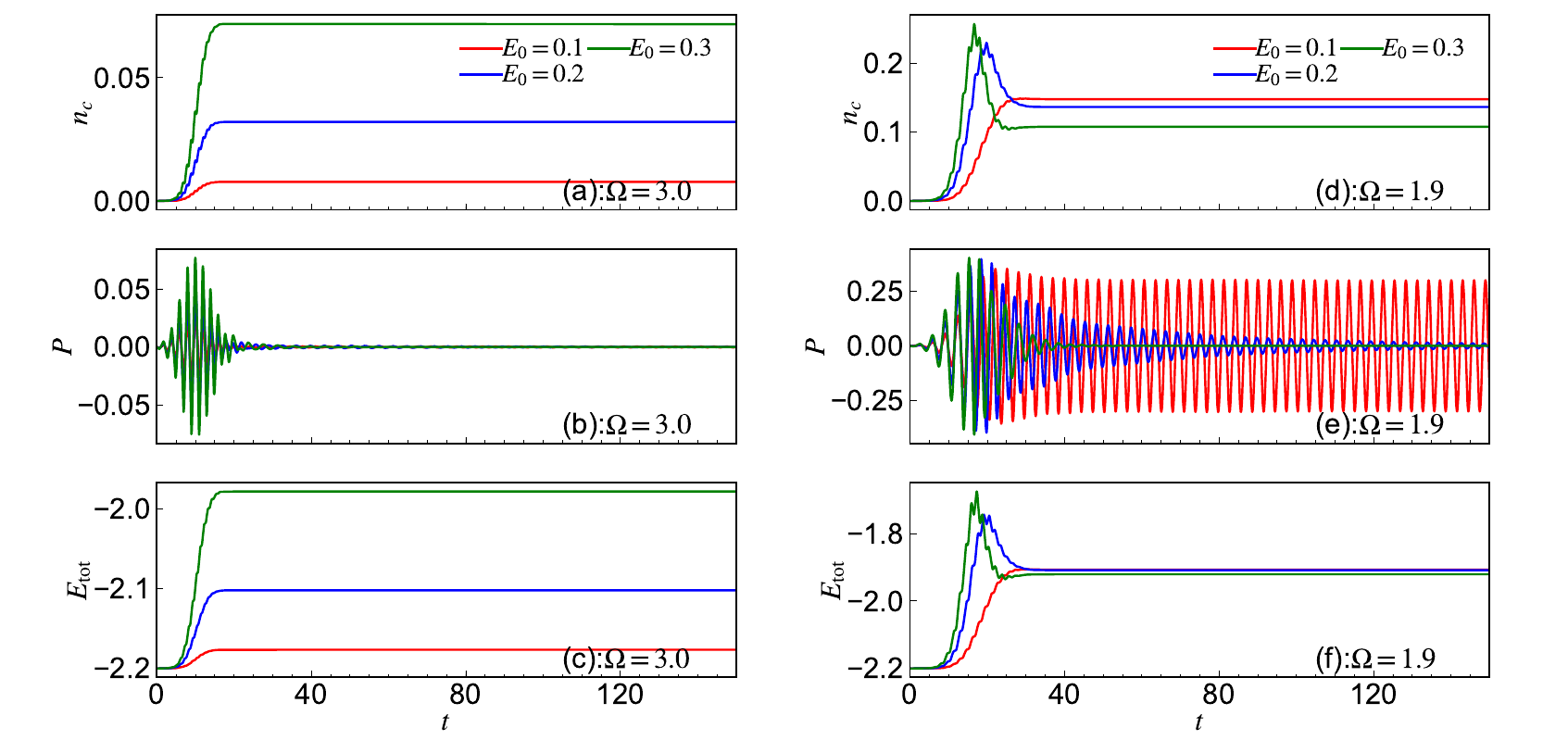} 
  \caption{GKBA+s2BA time evolution of the excited charge (a),(d), the dipole moment (b),(e),  and the total energy (c),(f) during and after the photo excitation with $\Omega=3.0$ (a)-(c) and $\Omega=1.9$ (d)-(f), pulse parameters defined in Eq.~\eqref{eq:pulse_info}, and different field strengths $E_0$.}
  \label{fig:GKBA_U2Ome30}
\end{figure*}

\subsection{Beyond linear response} \label{sec:results_nonlinear}
Now, we discuss the time evolution of the system during and after a photo-excitation beyond the linear response regime.
In the following, we use $\Delta_v = -3.2$ and $\Delta_c=1.2$, and $U=2.0$, which gives ${\mathcal E}_{\rm gap}^{\rm eq}=2.4$ and ${\mathcal E}_{\rm ex}^{\rm eq}=1.93$ in equilibrium at $T=0$.
We apply the Gaussian pulse with
\begin{align}
A_x(t)&=A_0\cdot F_{\rm gauss} (t-t_0,\sigma)\cdot \sin(\Omega(t-t_0)) \cdot F_{\rm ramp}(t,t_r). 
\end{align}
Here 
$F_{\rm gauss} (t,\sigma) (= \exp\bigl(-\frac{t^2}{2\sigma^2}\bigl))$ is the envelope function and 
\begin{align}
F_{\rm ramp} (t,t_r) & =
\begin{cases}
0 \;\;\text{   for ($t\leq 0$)}\\
\frac{1}{2}-\frac{3}{4}\cos(\pi t/t_r)+\frac{1}{4}\cos(\pi t/t_r)^3 \\
\phantom{0}\;\;\text{   for ($0<t<t_r$)}\\
1 \;\;\text{   for ($t_r\leq t$)}
\end{cases}
\end{align}
is a ramp-up function which ensures that the evolution of the field around $t=0$ is smooth.
In the following, we use
\begin{align}
\phi=0,\;\; t_0 = \frac{N_{\rm cyc} \pi }{\Omega},\;\; \sigma=\frac{t_0}{3.0},\;\; t_r = \frac{2\pi}{8\Omega}, \label{eq:pulse_info}
\end{align}
with $N_{\rm cyc} = 10$ unless we mention the condition specifically.
Here $N_{\rm cyc}$ is the number of cycles included within $[-3\sigma,3\sigma]$ of the Gaussian envelope.
We will consider two cases, i) an excitation into the particle-hole continuum ($\Omega>{\mathcal E}_{\rm gap}^{\rm eq}$) and ii) a resonant excitation of the excitons ($\Omega = {\mathcal E}_{\rm ex}^{\rm eq}$).  The former case is depicted in Figs.~\ref{fig:dispersion_eq} and ~\ref{fig:VEffect_HF_Free_comparison}(c) with green arrows, while the latter is shown with blue arrows.
We note that in the case of strong excitations, ${\mathcal E}_{\rm ex}$ shifts away from 
its equilibrium value (${\mathcal E}_{\rm ex}^{\rm eq}$) during the pulse, so that for a fixed pulse frequency, the system eventually deviates from the resonant condition. 
With this excitation protocol, we are going to investigate how the exciton frequency ${\mathcal E}_{\rm ex}$, the binding energy ${\mathcal E}_{\rm b}$, and the single particle spectrum are affected by the photo-doping of the system. 

 \begin{figure*}
  \centering
    \hspace{-0.cm}
    \vspace{0.0cm}
   \includegraphics[width=170mm]{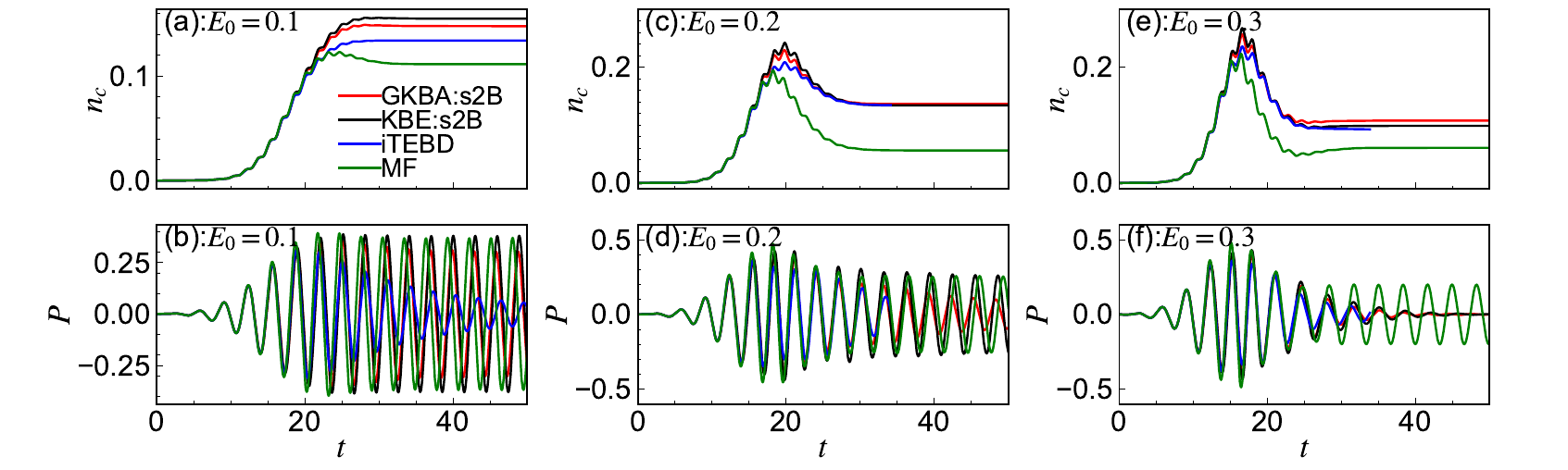} 
  \caption{Comparison of the density of conduction band electrons and the polarization among s2BA, GKBA+s2BA, iTEBD and MF for $\Omega=1.9$.
   (a),(b) are for $E_0=0.1$, (c),(d) are for $E_0=0.2$ and (e),(f) are for $E_0=0.3$.}
  \label{fig:Compare_U2Ome19E01_02}
\end{figure*}

In Fig.~\ref{fig:GKBA_U2Ome30} 
we first show the GKBA+s2BA results for the time evolution of the number of electrons in the conduction band, the dipole moment, and the total energy after different excitations.
For $\Omega=3.0>{\mathcal E}_{\rm gap}^{\rm eq}$ (left panels), the number of excited charge carriers increases with increasing field strength in this regime.
In the absence of  a field, the bands are decoupled and the Hamiltonian conserves the number of particles in the conduction and valence band, respectively, 
which is correctly captured by the GKBA.
As expected, since $\Omega$ is far from the exciton frequency, there is no prominent oscillation observed after the pulse, 
which lasts up to $t\approx 20$.
For $\Omega=1.9\simeq {\mathcal E}_{\rm ex}^{\rm eq}$ (right panels), one can observe a non-monotonic increase of $n_c$ as a function of time as well as the field strength. This can be understood as a Rabi oscillation between the ground state and the exciton state.
After the excitation ($t\gtrsim 30$), one can observe strong coherent oscillations in $P(t)$ with some frequency $\omega_{\rm coh}$, which persist for a long time after the pulse. 
The damping speed of these oscillations is enhanced with increasing field strength.   
From the Fourier transformation of these oscillations, one finds $\omega_{\rm coh}=2.13$ ($\omega_{\rm coh}=2.12$) for $E_0=0.1$ ($E_0=0.2$)
 at $t=60$ (The frequency is a bit $\sim0.04$ increased from just after the pulse.). These values {\it exceed} the exciton frequency in equilibrium and the renormalized gap energy ${\mathcal E}_{\rm gap}^\text{ren}=1.91$ (${\mathcal E}_{\rm gap}^\text{ren}=1.85$).
Here ${\mathcal E}^{\rm ren}_{\rm gap}$ is extracted from the instantaneous MF hamiltonian ${\bf h}_{\rm HF}[\brho](t)$. 
Note that when the amplitude of the polarization becomes small the contribution of the Fock term becomes negligible and ${\mathcal E}^{\rm ren}_{\rm gap}$ is mainly determined by the Hartree shift. 
(For smaller field amplitude $E_0$, the oscillation frequency is still smaller than the renormalized gap energy.)
As demonstrated in Figs.~\ref{fig:GKBA_U2Ome30}(c) and \ref{fig:GKBA_U2Ome30}(f), the total energy ($E_{\rm tot}$) is conserved after the excitation.

Now let us compare the results obtained by the different numerical methods. 
In Fig.~\ref{fig:Compare_U2Ome19E01_02},
we compare the density of the conduction band electrons and the polarization obtained by s2BA, GKBA+s2BA, MF and iTEBD for $\Omega=1.9$.
In all cases, the strong coherent oscillations in the polarization persist even after the pulse.
Among the approximate methods (s2B, GKBA+s2B, MF), GKBA provides the results closest to those of iTEBD.
The most important difference between the tdMF and the rest is the damping of the induced coherent oscillations.
Although GKBA still underestimates the damping compared to iTEBD, 
we find that the estimation of the damping within the GKBA is quantitatively better for the stronger fields.
The s2BA can also show the damping of oscillations but it is generally weaker compared to GKBA and for $E_0=0.1,0.2$ it is very weak, while 2BA and GKBA match better as we further increase the field strength.
Importantly, the peculiar feature of the coherent oscillations induced by the resonant excitation can be observed in iTEBD. 
For example, within iTEBD $\omega_{\rm coh}$  is $2.05$, while ${\mathcal E}_{\rm gap}$ is $1.86$ for $E_0=0.1$ around $t=60$.
(Since the direct evaluation of the single particle gap in nonequilibrium iTEBD calculations is difficult, we estimate ${\mathcal E}_{\rm gap}$ from the density of excited charges $n_c$ and the resulting Hartree shift.)
We also compare 2BA, GKBA+2BA, s2BA and GKBA+s2BA in Appendix~\ref{appendixC}, but, in the present setup, the exchange term does not result in a significant change in the evolution of $P$
nor systematically improve  the results compared to s2BA and GKBA+s2BA.
This comparison suggests that the GKBA captures well the relevant features of the dynamics of the extended systems and that it is a useful method for systematic studies due to its  relatively cheap computational cost.

Let us now comment on the relation between the strong coherent oscillations observed here and results reported in previous works.\cite{Comte1986,Ostreich1993,Perfetto2019PRM,Perfetto2019PRM,Littlewood2001PRB,Littlewood2006PRL}
After the excitation, the Hamiltonian conserves the number of electrons and holes, respectively.
If the excited charge carriers are cooled down due to some coupling to thermal baths, the steady state after the relaxation 
should be described by a thermal state of the original Hamiltonian (Eq.~(1) without excitation) with two \emph{different} effective chemical potentials for the  conduction band ($\mu_c$) and valence band ($\mu_v$), $\hat{H}^M =\hat{H}(0) -\mu_c \hat{N}_c -\mu_v \hat{N}_v$.\cite{Perfetto2019PRM,Littlewood2001PRB,Littlewood2006PRL} 
Here $\mu_c$ and $\mu_v$ are determined such that the number of electrons and holes
is the same as that just after the excitation.
Since $\hat{H}^M$ corresponds to the original Hamiltonian with a smaller band gap, it can exhibit  
an excitonic insulating (EI) phase (exciton condensation out of equilibrium).\cite{Perfetto2019PRM,Littlewood2006PRL} 
The time evolution of the system is however described by $\hat{H}$, not by $\hat{H}^M$, so that this state exhibits oscillations of the polarization (off-diagonal component of the density matrix)
with frequency $|\mu_c-\mu_v|$, which is of the order of the band gap.
It has recently been shown in an independent work (Ref.~\onlinecite{Perfetto2019PRM}) using the same model as considered here and tdMF, that such a state can be realized even without baths by using a suitable pulse shape.
Thus, the strong coherent oscillations in the polarization observed here can be understood as a transient realization of a nonequilibrium EI phase and the frequency of the oscillations above the renormalized band gap can be attributed to the effective chemical potentials in the two bands.
Still, we have to note that the state just after the excitation is not exactly the thermal equilibrium state of $\hat{H}^M$, and our beyond-MF simulation shows that the transient EI phase can decay because of the scattering between the excited carriers. In the following, we focus on the properties of the transient states characterized by a gradually decreasing polarization. 
The effects of a coupling to phonon baths, which results in the realization of an equilibrium state of $\hat{H}^M$, are discussed in Sec.~\ref{sec:el_ph_effects}.
 \begin{figure*}[t]
  \centering
    \hspace{-0.cm}
    \vspace{0.0cm}
     \includegraphics[width=85mm]{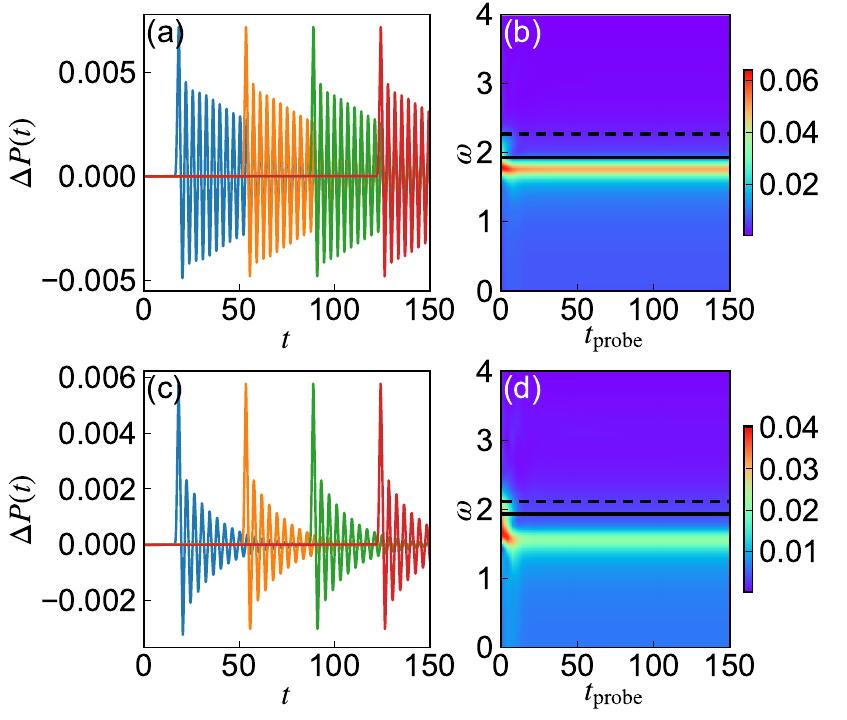} 
   \includegraphics[width=85mm]{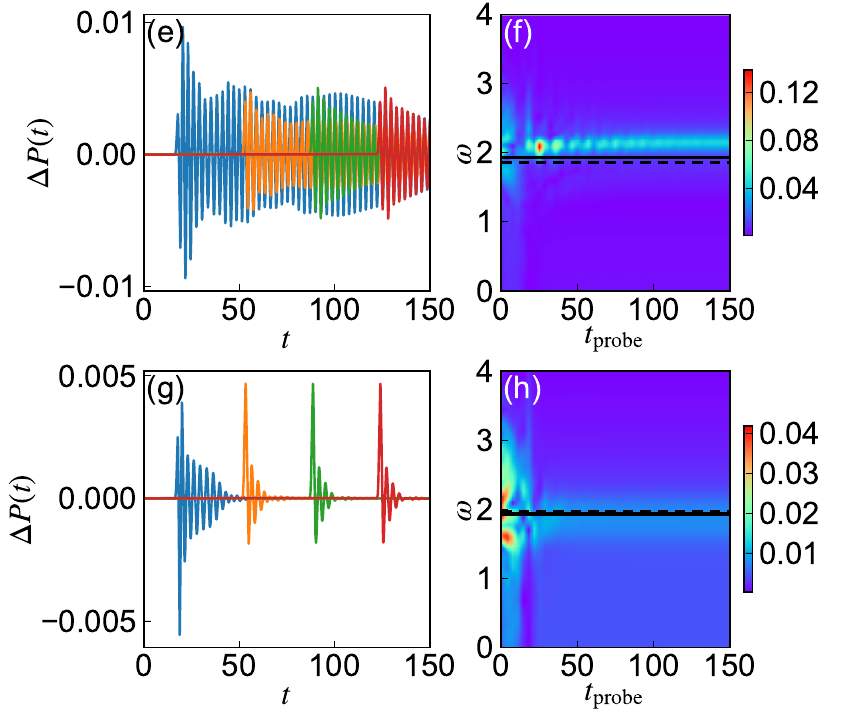} 
  \caption{ (a), (c), (e), (g) The difference in the dipole moment $\Delta P (t;t_{\rm probe}$) between the cases with and without the probe pulse for different delay times. (b), (d), (f), (h) The Fourier transformation of $\Delta P$ with respect to $t$ ($|\Delta P (\omega;t_{\rm probe})|$ defined in Eq.~\eqref{eq:DP_w}) is plotted in the space of $\omega$ and $t_{\rm probe}$. Panels (a)-(d) show the result for pump pulse frequency $\Omega=3.0$ and (e)-(h) for $\Omega=1.9$. The other pulse parameters are defined in Eq.~\eqref{eq:pulse_info}, and 
  the field strength of the pump pulse is $E_0=0.2$ and $E_0=0.3$ for (a), (b), (e),(f) and (c), (d), (g), (h) respectively. Black solid lines indicate ${\mathcal E}_{\rm ex}^{\rm eq}$ and back dashed lines show the renormalized band gap $\mathcal{E}_{\rm gap}^{\rm ren}$ after the excitation.
  }
  \label{fig:GKBA_pp_Ome30_19}
\end{figure*}
 \begin{figure}[t]
  \centering
    \hspace{-0.cm}
    \vspace{0.0cm}
   \includegraphics[width=85mm]{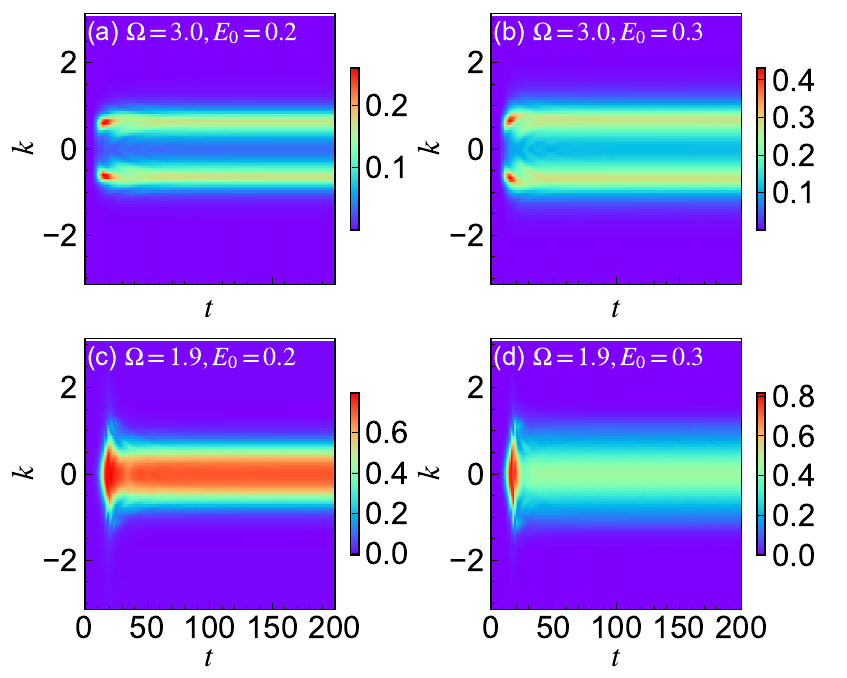} 
  \caption{GKBA + s2BA time evolution of the momentum distribution of the conduction-band electrons ($n_c(k)$) for different pump pulse excitations and amplitudes. }
  \label{fig:nk_GKBA}
\end{figure}

To study properties of the transient states, we perform a pump-probe simulation using GKBA + s2BA.
Namely, in addition to the first strong pump field, we add a second weak pulse (probe pulse) with some time delay.
The shape of the probe pulse is chosen as 
\begin{align}
E_{\rm probe}(t;t_{\rm probe}) = E_{\rm probe} F_{\rm gauss}(t-t_{\rm probe},\sigma_{\rm probe}).
\end{align}
In the following we use $\sigma_{\rm probe}=0.5$ and $E_{\rm probe}=0.01$ and neglect the vector potential of the probe pulse.
Then we measure the dipole moment $P(t)$ and calculate the difference between the results with and without a probe pulse at $t_{\rm probe}$,
\begin{align}
\Delta P(t;t_{\rm probe}) \equiv P(t;t_{\rm probe}) - P(t)_\text{no probe}.
\end{align}
To identify frequencies of oscillations induced by the probe pulse at $t_{\rm probe}$, we perform a Fourier transformation with a window function,
\begin{align}
\Delta P(\omega;t_{\rm probe}) = \int dt \;\; \Delta P(t;t_{\rm probe}) F_{\rm window}(t;t_{\rm probe}) e^{i\omega t}. \label{eq:DP_w}
\end{align}
Here $F_{\rm window}(t;t_{\rm probe})=F_{\rm gauss}(t-t_{\rm probe},\sigma)$ and $\sigma=20.0$ is used in the following.
This time dependent spectral function can reveal the excitation structure of the transient state around $t=t_{\rm probe}$, when 
the oscillations induced by the pump pulse is not large or slower than the characteristic frequency induced by the probe pulse.
We call the peak in $\Delta P(\omega;t_{\rm probe})$ as $\omega^*_{\rm coh}$ in the following.

In Fig.~\ref{fig:GKBA_pp_Ome30_19} 
we show the results of these analyses for $\Omega=3.0$ and 
$\Omega=1.9$, respectively. For $\Omega=3.0$ (above band-gap excitation, left four panels), one finds that there is almost no change in $\Delta P(t;t_{\rm probe})$ and hence $\Delta P(\omega;t_{\rm probe})$ after the pump pulse. With increasing field strength, 
the oscillation frequency ($\omega^*_{\rm coh}$) becomes smaller and at the same time, the life time of the oscillation becomes shorter.
After the excitation, the band gap is reduced because of the Hartree shift from the photo carriers.
Still, the frequency of the oscillation is within the shifted band gap, and thus the situation is qualitatively similar to the exciton state in equilibrium.
Thus we can regard $\omega^*_{\rm coh}$ as a renormalized exciton energy, $\mathcal{E}^{\rm ren}_{\rm ex}$.
We note that within GKBA + s2BA, the renormalized binding energy, $\mathcal{E}_{\rm b} =\mathcal{E}^{\rm ren}_{\rm gap} - \mathcal{E}^{\rm ren}_{\rm ex}$, is slightly increased to $0.50$ ($0.55$) for $E_0=0.2$ ($E_0=0.3$) from the equilibrium value $\mathcal{E}_{\rm b,eq}=0.47$. 
However, within GKBA + 2BA, even though $\mathcal{E}^{\rm ren}_{\rm ex}(=\omega^*_{\rm coh})<\mathcal{E}^{\rm ren}_{\rm gap}$, $\mathcal{E}_{\rm b}=0.42$ for $E_0=0.2$. Whether the enhancement of $\mathcal{E}^{\rm ren}_{\rm b}$ is genuine or not is thus unclear.  (iTEBD can only access short times for $\Omega=3.0$.)

For $\Omega=1.9$ (resonant excitation, right four panels), one observes a very different behavior from the case discussed above. 
Namely, the frequency of the induced oscillations ($\omega^*_{\rm coh}$) increases for small $E_0$ from ${\mathcal E}_{\rm ex}^{\rm eq}$ and decreases for larger $E_0$.
More remarkably, the frequency can {\it exceed} the renormalized band gap unlike the normal exciton in equilibrium.
We note that, when the nonequilibrium states induced by the pump pulse show strong oscillations, the signal induced by the probe pulse also follows these oscillations and $\omega^*_{\rm coh}$ becomes similar to $\omega_{\rm coh}$. 
Hence, the gradual shift of $\omega^*_{\rm coh}$ after the pulse for $E_0=0.2$ can be attributed to the shift of $\omega_{\rm coh}$ itself. 
When the amplitude of the oscillations induced by the pump pulse is damped and becomes small, $\omega_{\rm coh}$ and $\omega^*_{\rm coh}$ are essentially the same, since both oscillations can be regarded as a small perturbation around the state without the oscillations.
As in the case of $\Omega=3.0$, the life-time of the oscillations becomes shorter with increasing field strength.

Since the exciton states should be strongly affected by the transient quasiparticle occupations,
we study the time evolution of the momentum distribution of the charges ($n_c(k),n_v(k)$).
Since $n_c(k)$ and $1-n_v(k)$ behave identically, we only show $n_c(k)$ in Fig.~\ref{fig:nk_GKBA}.
For $\Omega=3.0$ [Figs.~\ref{fig:nk_GKBA}(a) and \ref{fig:nk_GKBA}(b)], the charges are excited at finite momenta which correspond to $\Omega = E_{c}(k)-E_{v}(k)$. 
Even though there occurs a slight redistribution and the occupation around $k=0$ becomes nonzero, most of the excited charges remain at nonzero momentum, and after the pulse the nonthermal distribution function remains almost constant. 
This is qualitatively similar to the MF dynamics, even though the latter lacks scattering and the occupation around $k=0$ remains almost zero after the pulse, see Fig.~\ref{fig:nk_MF} in Appendix~\ref{appendixB}.
The slow intra-band relaxation is a consequence of the one-dimensional setup we are using, which implies that the scattering between charges is strongly restricted because of the momentum conservation and the energy conservation.
It is expected that if we use a higher-dimensional lattice or consider electron-phonon scattering, 
one can observe a faster thermalization/redistribution process. In Sec~\ref{sec:el_ph_effects} we will analyze the effects of  electron-phonon couplings.

For $\Omega=1.9$ [Figs.~\ref{fig:nk_GKBA}(c) and \ref{fig:nk_GKBA}(d)], the charges are directly excited around $k=0$. After the pulse, the distribution function remains almost unchanged.
The comparison with the results from tdMF (Fig.~\ref{fig:nk_MF} in Appendix~\ref{appendixB}) shows that the redistribution of the population due to 
scattering is indeed captured by GKBA, which yields a smooth distribution as a function of momentum and leads to an increase of the occupation near $k=0$.
In both simulations, a fast approach to a steady value is observed after the pump, which is consistent with a change of the oscillation frequency during or quickly after the excitation.
We note that for  $E_0=0.3$  the particles are broadly distributed in the momentum space compared to the case for $E_0=0.2$.

To understand the origin of the qualitatively different collective modes ($\omega^*_{\rm coh}$) in the transient state after the pump pulse, depending on the excitation frequency, we now perform an RPA-type analysis using the essentially steady momentum distribution after the pulse. The idea of this analysis is the following. 
First, we extract, the momentum distribution $n_c(k)$ and $n_v(k)$ after the pulse from the GKBA simulation.
We then substitute these  $n_c(k)$ and $n_v(k)$ (neglecting the interorbital components $\langle \hc^\dagger_c \hc_v \rangle, \langle \hc^\dagger_v \hc_c \rangle$) into Eqs.~\eqref{eq:chi0_noneq_w}, \eqref{eq:gamma022_noneq_w} and \eqref{eq:rpa_like} to estimate the nonequilibrium susceptibility. 
We note that this approximation corresponds to the MF dynamics starting from the distribution given by  $n_c(k)$ and $n_v(k)$ (without interorbital component), which is a steady-state solution of the MF equation of motion.

 \begin{figure}[t]
  \centering
    \hspace{-0.cm}
    \vspace{0.0cm}
   \includegraphics[width=85mm]{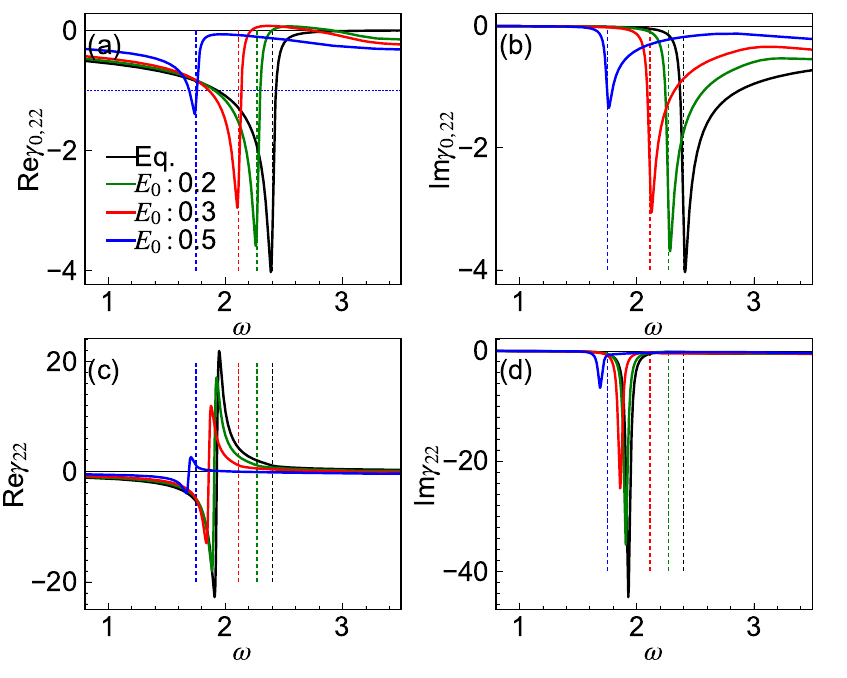} 
  \caption{Results of $\gamma_{0,22}$ [Eq.~\eqref{eq:gamma022_noneq_w}] and $\gamma_{22}$ [Eq.~\eqref{eq:gamma_rpa}] for different field strengths $E_0$ and $\Omega=3.0$. The momentum distribution is obtained from the GKBA +s2BA simulation at $t=150$. The vertical dotted lines indicate the band gap estimated by the MF Hamiltonian, Eq.~\eqref{eq:H_MF2}. Here we set $0^+ = 0.02$ in Eq.~\eqref{eq:gamma_noneq_w} for the numerical evaluation, which explains the finite 
  weight in Im$\gamma_{0,22}$ below the band gap.  "Eq." indicates the results in equilibrium.}
  \label{fig:Noneq_RPA_Ome30}
\end{figure}

 \begin{figure}[th]
  \centering
    \hspace{-0.cm}
    \vspace{0.0cm}
   \includegraphics[width=85mm]{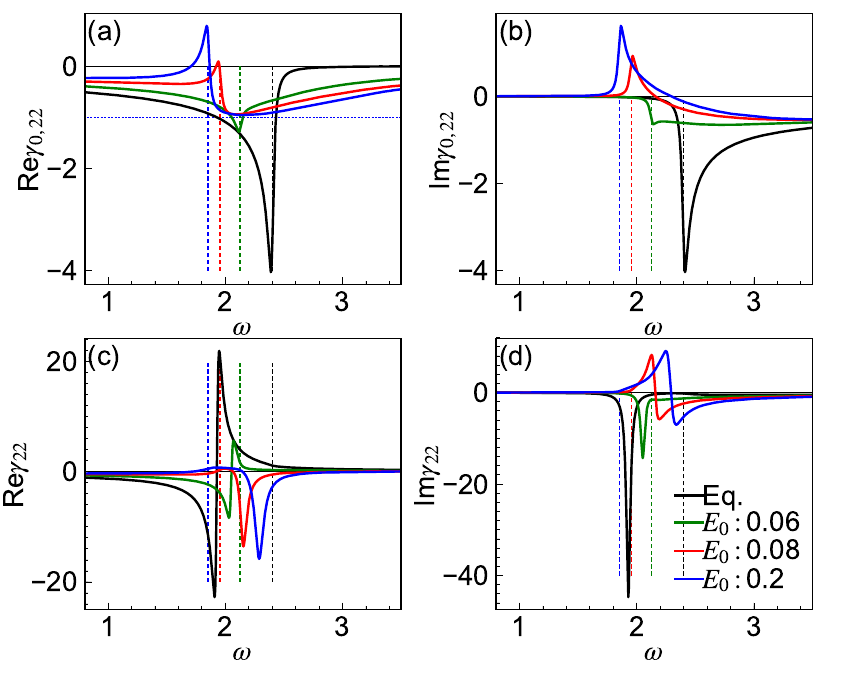} 
  \caption{Results of $\gamma_{0,22}$ (Eq.~\eqref{eq:gamma022_noneq_w}) and $\gamma_{22}$ (Eq.~\eqref{eq:gamma_rpa}) for different field strengths and $\Omega=1.9$. The momentum distribution is obtained from the GKBA + s2BA simulation at $t=150$. The vertical dotted lines indicate the band gap estimated by the MF Hamiltonian, Eq.~\eqref{eq:H_MF2}. Here we set $0^+ = 0.02$ in Eq.~\eqref{eq:gamma_noneq_w} for the numerical evaluation, which explains the finite 
  weight in Im$\gamma_{0,22}$ below the band gap. "Eq." indicates the results in equilibrium.}
  \label{fig:Noneq_RPA_Ome19}
\end{figure}
In Figs.~\ref{fig:Noneq_RPA_Ome30} and ~\ref{fig:Noneq_RPA_Ome19}, we show $\gamma_{0,22} (\omega)$ and $\gamma_{22} (\omega)$ for different pump frequencies and amplitudes. 
For $\Omega=3.0$ (Fig.~\ref{fig:Noneq_RPA_Ome30}), as we increase the field strength, more electrons are excited to the conduction band and the band gap becomes smaller because of the Hartree shift, see Eq.~\eqref{eq:Hartree}.
As a consequence, the edge of the imaginary part of $\gamma_{0,22}$ is shifted to lower energies and the peak at the edge is reduced because of the finite density of conduction electrons around $k=0$, see Eq.~(\ref{eq:gamma022_noneq_w}).
The electrons stuck at non-zero momentum appear in the imaginary part of $\gamma_{0,22}$ as a local minimum around $\omega=3.0$. Because the imaginary part of $\gamma_{0,22}$ is connected to the real part through the Kramers-Kronig relation, these features in the imaginary part lead to a shift of the peak and a reduction of the height of the peak in the real part.
Still the peak in the real part in $\gamma_{0,22}$ is prominent, which leads to a well defined in-gap mode appearing in 
the imaginary part of $\gamma_{22}$, see Fig.~\ref{fig:Noneq_RPA_Ome30}(d).
The exciton binding energy (the distance between the peak and the dashed line in Fig.~\ref{fig:Noneq_RPA_Ome30}(d)) is gradually reduced with increasing pulse amplitude, which reflects the reduction of the height of the peak in the real part of $\gamma_{0,22}$.

 \begin{figure*}
  \centering
    \hspace{-0.cm}
    \vspace{0.0cm}
   \includegraphics[width=170mm]{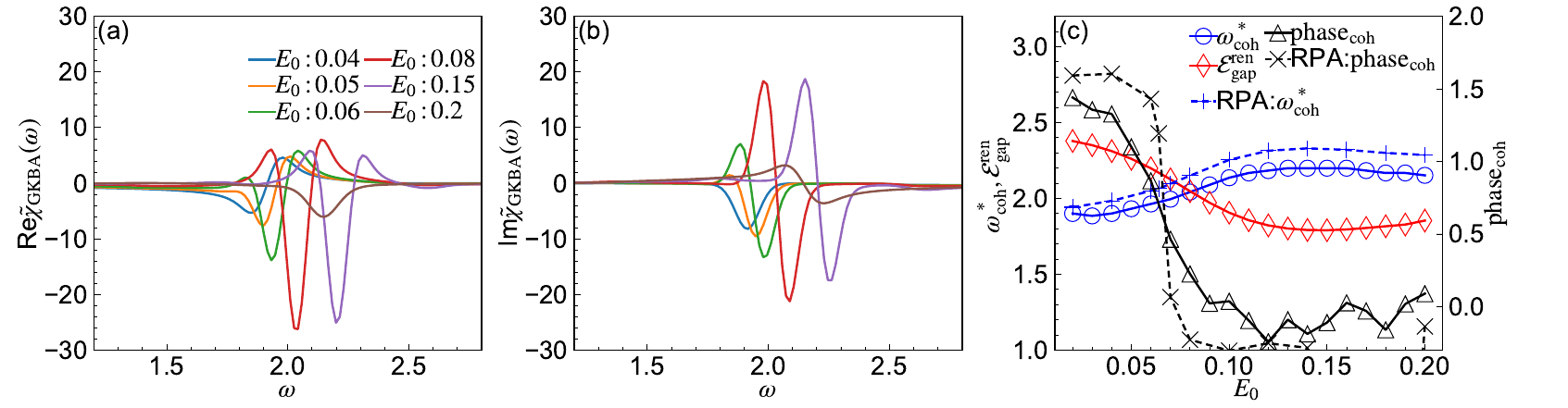} 
  \caption{(a), (b) Real and imaginary part of the  susceptibility estimated by the GKBA + s2BA simulation averaged around $t_{\rm probe}=150$ ($\tilde \chi_{\rm GKBA} (\omega)$). See the main text for detailed explanations.  (c) Summary of the frequency of the oscillation ($\omega^*_{\rm coh}$), the band gap and the phase of the susceptibility at $\omega = \omega^*_{\rm coh}$. 
  Here, $\omega^*_{\rm coh}$ is estimated by the peak position in $|\tilde \chi_{\rm GKBA} (\omega)|$ or $\chi_{11}(\omega)$ and the phase is defined as the argument of $-\tilde \chi_{\rm GKBA} (\omega^*_{\rm coh})$ or  $-\chi_{11}(\omega^*_{\rm coh})$. We note that the gap size is estimated by the GKBA analysis at $t=150$.
  }
  \label{fig:GKBA_summary}
\end{figure*}

For $\Omega=1.9$ (Fig.~\ref{fig:Noneq_RPA_Ome19}), we observe a suppression of the band gap with increasing field strength.
Different from the case of $\Omega=3.0$, the excited charges directly accumulate at the bottom of the conduction band 
around $k=0$. 
This produces a more drastic change in $\gamma_{0,22}$ and hence in $\gamma_{22}$.
For $E_0=0.06$, the peak structure around the renormalized gap ${\mathcal E}^{\rm ren}_{\rm gap}$ is strongly suppressed in $\gamma_{0,22}(\omega)$, but there still exists a crossing 
between the real part of $\gamma_{0,22}(\omega)$ and $-2/U$, which leads to a well-defined in-gap state as in equilibrium, see Fig.~\ref{fig:Noneq_RPA_Ome19}(c,d).
When we further increase $E_0$, a population inversion ($n_c(k)>n_v(k)$) occurs around $k=0$, which is reflected in
the positive value of Im$\gamma_{0,22}(\omega)$ around ${\mathcal E}^{\rm ren}_{\rm gap}$. 
Because of this population inversion near $k=0$, the imaginary part of $\gamma_{0,22}(\omega)$ crosses zero at a certain 
energy, which we denote by $\omega^*$.
This zero-crossing can lead to a peak in the real part of  $\gamma_{22}(\omega)$.
To show this let us approximate $\gamma_{0,22}(\omega)\simeq \alpha + i\beta(\omega-\omega^*)$ around $\omega^*$. 
Using Eq.~\eqref{eq:gamma_rpa},
\begin{align}
\gamma_{22}(\omega) \simeq \frac{[\alpha + i\beta(\omega-\omega^*)][(1+\frac{U}{2}\alpha)-i\frac{U}{2}\beta(\omega-\omega^*)]}
{(1+\frac{U}{2}\alpha)^2+[\frac{U}{2}\beta(\omega-\omega^*)]^2}.
\end{align}
This expression features a pole at $\omega = \omega^* +i(1+\frac{U}{2}\alpha)/(\frac{U}{2}\beta)$.
If $(1+\frac{U}{2}\alpha)/(\frac{U}{2}\beta)$ is small compared to $\omega^*$ and the range in which the linearization of $\gamma_{0,22}(\omega)$ is justified, one can see a clear peak in the real part of $\gamma_{22}(\omega)$ around $\omega^*$.
This condition is indeed satisfied in the present case, see $E_0=0.08,0.2$ in Figs.~\ref{fig:Noneq_RPA_Ome19}(a)(b),
where $\alpha\gtrsim -2/U$, so that we end up with a clear peak in the real part of $\gamma_{0}(\omega)$, see Figs.~\ref{fig:Noneq_RPA_Ome19}(c)(d).

Thus, the RPA-type analysis qualitatively reproduces the dependence of the frequency of the collective mode $\omega^*_{\rm coh}$ on the excitation condition.
For the above-gap excitation with $\Omega=3.0$, $\omega^*_{\rm coh}$ stays smaller than $\mathcal{E}^{\rm ren}_{\rm gap}$, whose character is similar to that of excitons in equilibrium.
On the other hand, the mode observed for $\Omega=1.9$ above the renormalized band gap is explained by the population inversion just at the bottom of band (large $\beta$) and the moderate excitation, which results in a minimum of the real part of $\gamma_{0,22}(\omega)$ close to $-2/U$.
Since a resonant excitation at the equilibrium exciton energy can quickly induce such populations, 
its naturally result in
the peculiar coherent mode with frequency $\omega^*$.
Furthermore, the RPA-type analysis predicts that the appearance of a well-defined peak in the real part of the susceptibility $\boldsymbol{\chi}_0(\omega;{\bf q=0})$ instead of the imaginary part leads to a phase shift of the oscillation against the probe pulse.

We now directly check this change in the transient susceptibility within GKBA + s2BA.
Using GKBA, we can estimate the transient susceptibility through the pump-probe simulation as 
\begin{align}
\chi_{\rm GKBA}(\omega;t_{\rm probe})  = \frac{\Delta P(\omega;t_{\rm probe})}{-q\; E_{\rm probe}(\omega;t_{\rm probe})}
\end{align}
with $\Delta P(\omega;t_{\rm probe})$ defined in Eq.~\eqref{eq:DP_w} and $E_{\rm probe}(\omega;t_{\rm probe}) = \int dt e^{i\omega t}E_{\rm probe}(t;t_{\rm probe})$.
We note that this corresponds to $\chi_{11}^R(\omega;{\bf q=0})$ in equilibrium when $E_{\rm probe}$ is very weak.
Considering the fact that the system is oscillating, we calculate the average of $\chi_{\rm GKBA}(\omega;t_{\rm probe})$ over the time interval $145 \le t_{\rm probe} \le 155$ ($\tilde{\chi}_{\rm GKBA}(\omega)$)
and show the results in Fig.~\ref{fig:GKBA_summary}(a)(b).

For small $E_0$, there is a peak in the imaginary part of $\tilde{\chi}_{\rm GKBA}$, see $E_0=0.04$ as an example, while for large enough $E_0$, the peak appears in the real part of $\tilde{\chi}_{\rm GKBA}$, see e. g. $E_0=0.2$. This is consistent with the RPA-type analysis.
In Fig.~\ref{fig:GKBA_summary}(c), we show the renormalized gap (evaluated at $t=150$) and the frequency of the induced oscillation $\omega^*_{\rm coh}$ evaluated by the peak position of 
$|\tilde{\chi}_{\rm GKBA}(\omega)|$ as a function of $E_0$.
The relative magnitude of these quantities switches around $E_0=0.08$ but the weak- and strong-field regimes are smoothly connected (no singular behavior).
In Fig.~\ref{fig:GKBA_summary}(c), we also show the phase of  $-\tilde{\chi}_{\rm GKBA}(\omega)$ at $\omega = \omega^*_{\rm coh}$.
Reflecting a peak in the imaginary part for small $E_0$ and the one in the real part for large $E_0$, the phase quickly changes from a value close to $1.5$ to one close to $0$ near $E_0=0.08$. 

Although the GKBA and the RPA-type analyses agree qualitatively, there are several differences between them.
First, compared to GKBA, the RPA-type analysis shows a larger frequency of the collective mode and a more abrupt switching of the phase, Fig.~\ref{fig:GKBA_summary}(c).
Second, GKBA predicts that the signal in the crossover region becomes larger and the peak becomes sharper compared to the result for larger values of $E_0$, 
which is opposite to the behavior found in the RPA-type analysis.
We also note that for $E_0=0.17,0.19$, the RPA-type analysis predicts a positive weight at the peak in Re$\gamma_{22}$, which originates from the fact that $\gamma_{0,22}(\omega)$ becomes smaller 
than $-2/U$ at  $\omega^*$. Hence, the phase of $-\tilde{\chi}_{\rm GKBA}(\omega^*_{\rm coh})$ takes a value near $-\pi$.
In addition, the RPA-type analysis predicts an infinite life-time of the in-gap states, while in the GKBA analysis these states can decay.

 \begin{figure}[t]
  \centering
    \hspace{-0.cm}
    \vspace{0.0cm}
   \includegraphics[width=85mm]{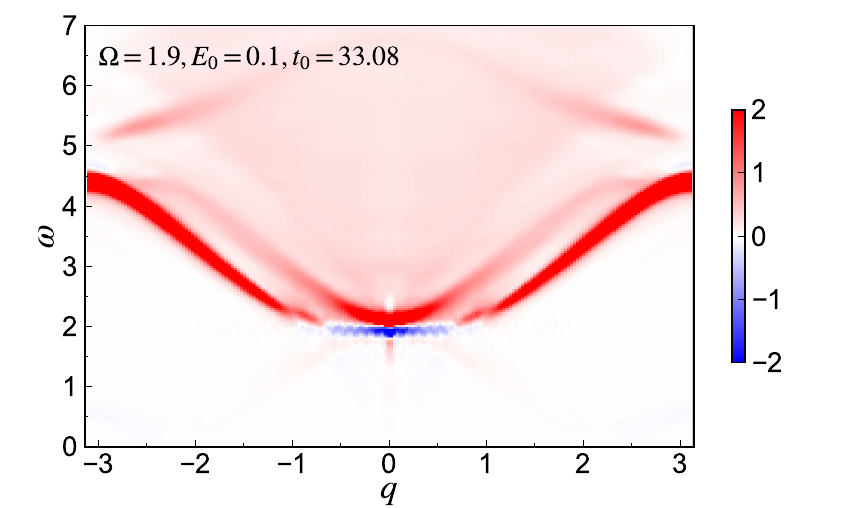} 
  \caption{ Imaginary part of the momentum resolved correlation function $-\mathrm{Im}\underline{\chi}^>(\omega;{\bf q},t_0)$ after the pump with $\Omega=1.9$ and $E_0=0.1$ for $t_0=33.08$ and $t_1=80$. Here $\sigma=(t_1-t_0)/(2\sqrt{2})$ is used for the window function. }
  \label{fig:chi_iTEBD}
\end{figure}

These differences may be attributed to i) the absence of the effects of the interorbital components in the RPA-type analysis, ii) the fact that GKBA partially takes into account  
the finite life-time of the quasiparticles as well as the corrections beyond the ladder diagrams from the correlated part of the self-energy.
Neglecting the effects of the off-diagonal part in the density matrix should not be justified 
when the induced oscillations are long-lived as in $E_0=0.06\sim 0.1$ at $\Omega=1.9$. 
Hence, the transition from the normal-exciton like oscillation to the peculiar collective mode above the band gap is not fully 
captured within the RPA-type analysis.
As for ii), the finite lifetime of quasi-particles can lead to a decay of the excitons and hence a finite lifetime, while  
the vertex corrections for the response functions beyond the RPA-type diagrams can renormalize the frequency of the oscillations.

Finally, we show the momentum resolved correlation functions evaluated by TEBD, Eq.~\eqref{eq:chi_iTEBD}.
In Fig.~\ref{fig:chi_iTEBD}, we show $-\mathrm{Im}\underline{\chi}^>(\omega;{\bf q},t_0)$  just after the resonant excitation ($\Omega=1.9$ and $E_0=0.1$), 
see Fig.~\ref{fig:VEffect_HF_Free_comparison}(c) for the equilibrium result.
In equilibrium, there is a single exciton band below the electron-hole continuum. 
The correlation function after the resonant excitation exhibits several well-defined bands.
Around $q=0$, the sign of $-\mathrm{Im}\underline{\chi}^>(\omega;{\bf q},t_0)$ changes around $\omega=2.05$, which is consistent with 
the behavior of $-{\rm Im}\tilde{\chi}^R_{\rm GKBA}$ when the peculiar mode is generated, see Fig.~\ref{fig:GKBA_summary}(b).
Interestingly, the positive signal above $2.05$ evolves into a well-defined branch at finite momentum which has a different dispersion than the exciton branch.

\subsection{Effects of electron-phonon coupling}\label{sec:el_ph_effects}

 \begin{figure}[t]
  \centering
    \hspace{-0.cm}
    \vspace{0.0cm}
   \includegraphics[width=85mm]{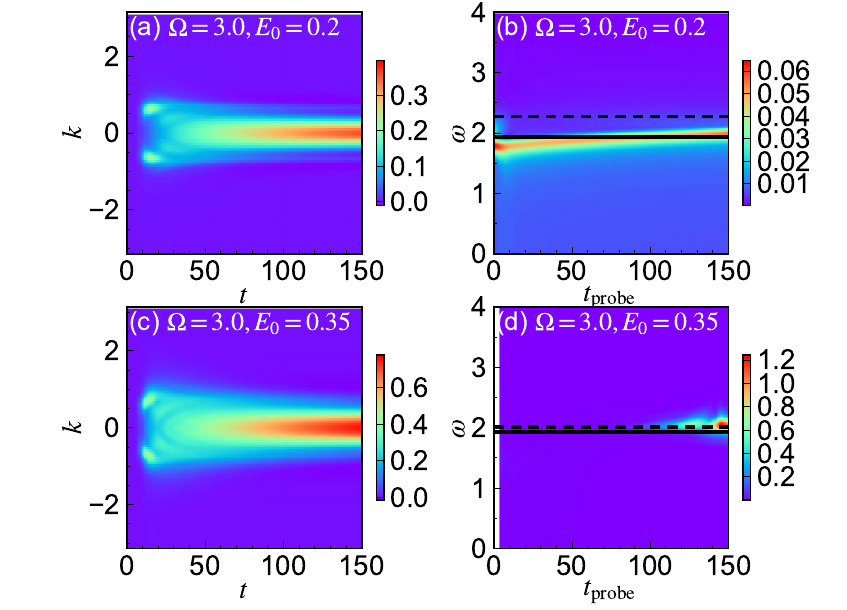} 
  \caption{(a), (c) Time evolution of the momentum distribution of the conduction-band electrons ($n_c(k)$) within GKBA + s2BA for finite electron-phonon couplings. 
  (b), (d) $|\Delta P (\omega;t_{\rm probe})|$ obtained by the pump-probe simulation (Eq.~\eqref{eq:DP_w})  plotted in the space of $\omega$ and $t_{\rm probe}$. 
  The solid black lines indicate the frequency of the exciton in equilibrium $\mathcal{E}_{\rm ex,eq}$, while the dashed black lines indicate the renormalized band gap $\mathcal{E}_{\rm gap}^{\rm ren}$, after the pulse measured at $t=150$. 
  (a), (b) is for $E_0=0.2$ and $\Omega=3.0$, while (c), (d) is for $E_0=0.35$ and $\Omega=3.0$.  Here $\omega_c=0.2$ 
  and $g=0.25$ are used.}
  \label{fig:el_ph}
\end{figure}

So far we have studied the dynamics of pure electron systems. 
However, in practice, there are nonzero electron-phonon (el-ph) couplings and the excited charge carriers can be cooled down. Especially in semiconductors, the relaxation in the conduction band can occur on a few tenth to a few hundred of femtoseconds and thus plays an import role.\cite{bar-ad_quantum_1997,banyai_exciton--lo-phonon_1995} The efficiency of the cooling depends on the strength of the el-ph coupling and the phonon frequency. Here we study the cooling effects using the GKBA.
Namely, in addition to the self-energy from the el-el interaction, Eq.~\eqref{eq:sigma2b_d}, we add the self-energy representing the el-ph coupling at the level of the Migdal approximation:
\begin{align}
\bSig^\gtrless_{\vec{k}}[\tilde{{\bG}}](t,t^\prime) = i g^2 \tilde{\bG}^\gtrless
_{\rm loc}(t,t^\prime) D^\gtrless_0(t,t^\prime) \ .
\end{align}
Here, $D_0(t,t^\prime)$ denotes the phonon GF and we assumed that the phonons are locally coupled to the densities of each band on each site.
Since the total density per site is fixed, with this type of coupling, no dynamics of the phonon displacement is induced after the excitation, and the Hartree-like (Ehrenfest) term can be ignored. Moreover, such a coupling to the phonon bath does not change the symmetry of the Hamiltonian so that the number of excited charges is conserved after the pulse.
We fix the phonon propagator to the equilibrium one (no feedback to the phonon subsystems), such that the phonons act as a heat bath.
The phonon GF is obtained by Fourier transforming $D^\gtrless_0(t,t^\prime) =  \int d\omega/(2\pi) D^\gtrless_0(\omega)e^{-i \omega(t-t^\prime)}$ and the fluctuation-dissipation theorem 
$D^>_0(\omega) = -i N_\mathrm{B}(\omega) B(\omega)$, $D^<_0(\omega) = -i [N_\mathrm{B}(\omega) + 1]B(\omega)$ ($N_\mathrm{B}(\omega)$ is the Bose distribution). Here we consider the Ohmic spectrum
\begin{align}
	B(\omega) = 2\pi \frac{\omega}{\omega_c} e^{-|\omega|/\omega_c}
\end{align}
with cutoff frequency $\omega_c$.

In Fig.~\ref{fig:el_ph}, we show the evolution of $n_{k,c}$ and the results of the pump-probe simulation for the excitation above the gap ($\Omega=3.0$) with different excitation strength.
One can see the relaxation of the excited carriers from finite momentum toward $k=0$, which was absent in the case without electron-phonon coupling.
Reflecting the time evolution of the momentum distribution, the frequency of the collective mode induced by the probe field gradually increases.
For the weaker excitation, the frequency of the collective mode remains below the band gap, while, for sufficiently strong excitations, at some point in time 
the frequency exceeds the renormalized band gap.
The latter result is very similar to the resonant excitation case without el-ph coupling, where the photo electrons (holes) are directly created at the bottom (top) 
of the conduction (valence) band. 
The present calculation shows that, with the cooling induced by the el-ph coupling and for sufficiently strong excitation,  
the peculiar collective mode can also be induced by above band-gap excitations. 

 \begin{figure}[t]
  \centering
    \hspace{-0.cm}
    \vspace{0.0cm}
   \includegraphics[width=85mm]{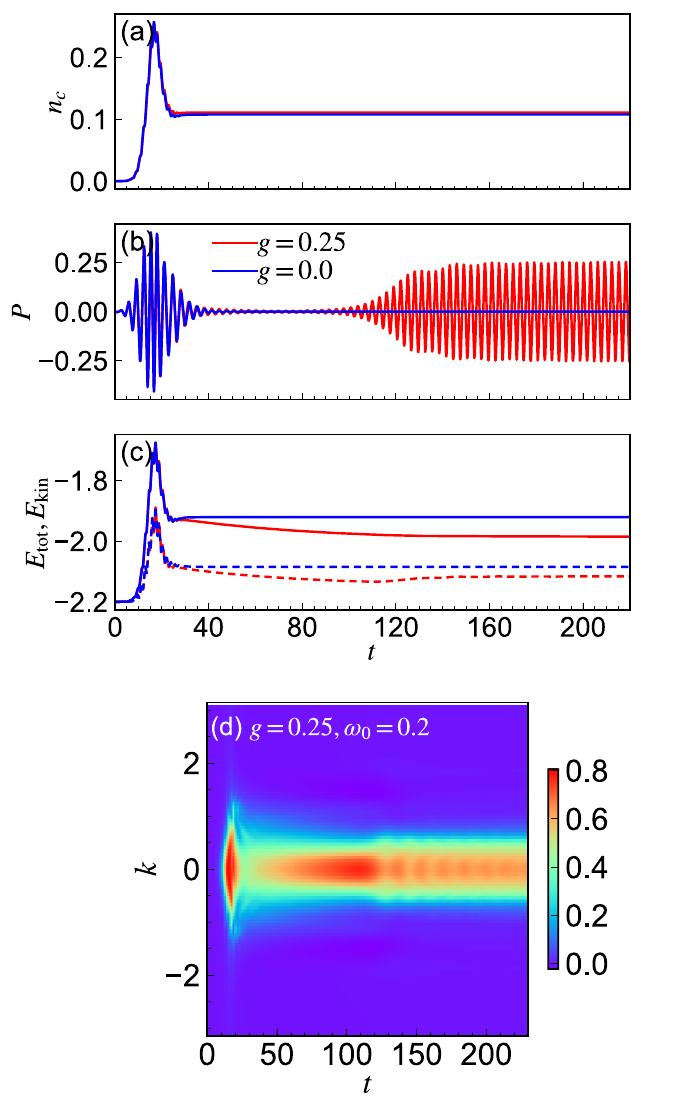} 
  \caption{(a)-(c) Comparison of the evolution after strong resonant excitation with and without phonon bath using GKBA+s2BA. Panel (a) shows the evolution of the excited charge, (b) shows the dipole moment, and (c) shows the total energy (solid line) and the kinetic energy (dashed line). (d) Evolution of the momentum distribution of the conduction band electrons using GKBA + s2BA.
 Here, $\omega_c=0.2$, $g=0.25,0.0$, $\Omega=1.9,E_0=0.3$. The other pulse parameters are defined in Eq.~\eqref{eq:pulse_info}.} 
  \label{fig:Ome19_elph}
\end{figure}

As discussed in Sec.~\ref{sec:results_nonlinear}, when the system is completely relaxed after the excitation, the steady state reached should be described by the original Hamiltonian (Eq.~(1) without excitation) with two \emph{different} chemical potentials for the  conduction band ($\mu_c$) and valence band ($\mu_v$), $\hat{H}^M =\hat{H}(0) -\mu_c \hat{N}_c -\mu_v \hat{N}_v$. \cite{Perfetto2019PRM,Littlewood2001PRB} 
Since the ground state of such a Hamiltonian can be an
excitonic insulating (EI) phase, 
one can expect the appearance of large amplitude persistent oscillations of the polarization (nonequilibrium EI phase).
In Sec.~\ref{sec:results_nonlinear}, we showed that the resonant excitation can create a transient state close to such an EI phase, consistent with the recent results in Ref.~\onlinecite{Perfetto2019PRM}.
Here we show that, in the presence of el-ph coupling, after the initial decay of the transient EI state, the nonequilibrium EI phase is recovered due to the cooling effect. As a result, large-amplitude persistent oscillations of the polarization reappear  at long times.

In Fig.~\ref{fig:Ome19_elph}, we compare the time evolution with and without the phonon bath for the strong resonant excitation, which generates the excited electrons near the  $\Gamma$ point. 
For the present field strength, the polarization damps quickly after the pulse in both cases.
However, in the presence of the phonon bath, the polarization recovers after some time and exhibits persistent oscillations,
which suggests the exciton condensation induced by the cooling of the excited charges.
In Fig.~\ref{fig:Ome19_elph}(c), we show the evolution of the kinetic and total energies. The phonon bath gradually reduces the total energy.
After the pulse, the kinetic energy also gradually decreases, but it starts to increase when the signal of the coherence of polarization starts to recover.
This is consistent with the Bardeen-Cooper-Schrieffer (BCS) scenario, since the ordered state lowers the interaction energy at the cost of increasing the kinetic energy.
In Fig.~\ref{fig:Ome19_elph}(d), we show the evolution of the momentum distribution of the conduction band electrons. 
One can clearly see that the electrons are more concentrated around the $\Gamma$ point compared to Fig.~\ref{fig:nk_GKBA}(d).
Slow oscillations in the density distributions set in around $t=120$, where the polarization starts to be enhanced.
These oscillations become less prominent in later times, which suggests the system approaches a steady state.

Finally, let us comment on a few points. First, we expect the emergence of the exciton condensation even in the case of the off-resonant excitation $\Omega=3.0$ if we simulate up to long enough times. 
Since the charges are excited to higher energy, it requires several scatterings with phonons for them to relax to the Gamma point. 
Second, a similar cooling and resultant condensation of excitons is expected for other types of el-ph couplings as long as the coupling does not break the symmetry. 
However, if the coupling is not to the total density on a given site,
one cannot ignore the phonon displacement and the resulting change of the electron energy levels due to the Ehrenfest term. 
Since the phonon displacement is expected to damp and approach some steady value, the steady state of the electrons will be determined by $\hat{H}^M$, taking into account the change of the energy levels due to the phonon displacement in a self-consistent manner.

\section{Summary and Conclusion}\label{sec:conclusion}

We have studied the fate of excitons in photo-excited semi-conductors using a spinless two band model in one dimension and different numerical methods; the tdMF, the 2BA, the GKBA implemented with 2BA and the iTEBD method. 
In the linear response regime at $T=0$, all these methods produce the exact linear response functions. Hence the exciton energies (${\mathcal E}_{\rm ex}$) 
can be accurately measured from the long-lived oscillations in the dipole moment after a weak excitation. 
Beyond the linear response regime, strong coherent oscillations in the polarization can be induced by resonant excitations.
 In particular, peculiar coherent oscillations characterized by a frequency larger than the semiconductor gap emerges for a properly chosen excitation strength.  
This behavior was also confirmed by the iTEBD simulation. We pointed out that these oscillations can be understood as a signature of the transient emergence of a nonequilibrium excitonic condensate,\cite{Ostreich1993,Perfetto2019PRM} which can be gradually suppressed as time evolves
by interaction effects beyond mean field.
Although 2BA and GKBA show 
such a suppression
they underestimate it compared to the iTEBD reference data. 
Still, GKBA captures relevant properties of the coherent oscillations and provides the best agreement with iTEBD among these approximate methods.

Focusing mainly on the GKBA results, we have closely analyzed the collective modes ($\omega^*_{\rm coh}$) of the transient stats after resonant and above-band-gap excitations
 using a numerical pump-probe simulation.
In the latter case, $\mathcal{E}_\text{ex}$(=$\omega^*_{\rm coh}$) is reduced mainly because of the photo-induced Hartree shift, but the exciton binding energy remains positive and thus the situation in the photo-doped state is qualitatively similar to an equilibrium state with reduced gap. 
For resonant excitations, $\omega^*_{\rm coh}$ tends to be increased from the equilibrium value of exciton $\mathcal{E}^{\rm eq}_\text{ex}$. 
When the excitation is weak, $\mathcal{E}^{\rm ren}_\text{gap}-\omega^*_{\rm coh}$ is still positive.
On the other hand, for stronger excitations, it can become negative but the mode induced by the probe pulse is still well-defined. 
We revealed the origin of this characteristic behavior using the RPA-type expression of the susceptibility and the nonequilibirum distributions from the GKBA analysis.
In particular, the peculiar mode characterized by the frequency above the band gap ($\omega^*_{\rm coh}>\mathcal{E}^{\rm ren}_\text{gap}$) originates from the photo-induced population inversion accompanied by a moderate number of excited charges and the sharp accumulation of electrons (holes) at the edge of the conduction (valence) band. The energy of this mode is determined by the energy 
up to which the photo-doped band is populated. 
We also studied the cooling effect from the electron-phonon coupling within GKBA.
Because of the cooling of excited carriers, the frequency of the collective mode evolves in time. 
We demonstrated that the efficient cooling of excited carriers and a sufficient amount of photo-doping can induce the peculiar mode above the band gap even after above-gap excitations.
We also simulated the build-up of the nonequilibrium exciton condensation in the phonon-cooled photo-doped state.

In the present study, we focused on a simplified model to benchmark the reliability of the methods and to explore potentially interesting phenomena.
Our study shows that GKBA essentially captures the relevant physics, which enables systematic analyses for extend systems at a reasonable computational cost.
In the future, it would be important and interesting to study the time evolution of excitons and charge distributions using more realistic models within GKBA. 
In addition, GKBA may be also be useful to study the real-time dynamics associated with the condensation of excitons or exciton polaritons out of equilibrium. The condensation problem has so far been mainly addressed with steady-state formalisms.
A more realistic model study would provide microscopic and detailed insights into the various nonequilibrium phenomena observed in transition metal chalcogenides as well as semiconductors in cavities.

\acknowledgements
YM, MS and PW were supported by ERC Consolidator Grant No. 724103 and the Swiss National Science Foundation via NCCR Marvel. The calculations were run on the Beo05 cluster at the University of Fribourg.  
M. S. thanks the Alexander von Humboldt Foundation for its support with a Feodor Lynen scholarship.

\appendix

\section{RPA-type analysis in the linear response regime} \label{sec:Tmat}
 \begin{figure}[th]
  \centering
    \hspace{-0.cm}
    \vspace{0.0cm}
   \includegraphics[width=85mm]{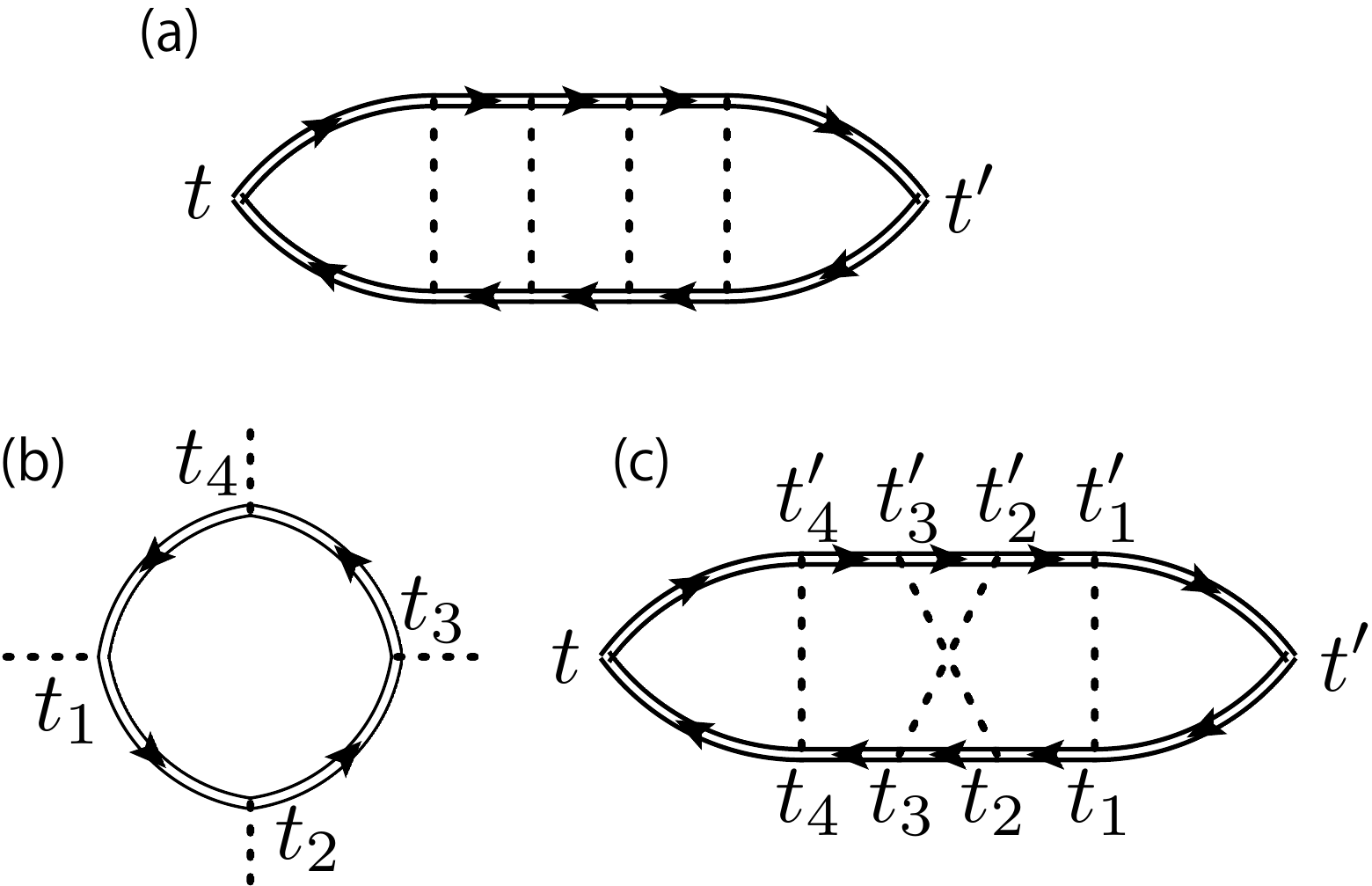} 
  \caption{(a) An example of ladder diagrams for $\chi_{\mu\nu}(t,t';{\bf q})$ in terms of the Feynman diagram. (b) an example of a ring contribution, which can appear in  the diagrams for  $\chi_{\mu\nu}(t,t';{\bf q})$. (c) An example of ladder-type diagrams with crossed interaction lines for  $\chi_{\mu\nu}(t,t';{\bf q})$.  Double lines with arrows indicates the full electron Green's function, while the dashed lines represent the Coulomb interaction.}
  \label{fig:diagrams_fig}
\end{figure}
For completeness, we provide a proof that the MF dynamics in the linear response regime is exact in the present model at $T=0$.
To this end, we consider the linear response in terms of the nonequilibrium Green's function (GF) formalism.\cite{Aoki2013,stefanucci_nonequilibrium_2013}
The electron GF is defined on the Konstantinov-Perel' contour (${\mathcal C}$) as in Eq.~\eqref{eq:def_gf}.
We also introduce the correlation function on the contour as 
\begin{align}
\chi_{\mu\nu}(t,t';{\bf q})& = -i\langle\mathcal{T}_{\mathcal C} \hat{\rho}_{\mu,{\bf q}}(t)  \hat{\rho}_{\nu,{\bf -q}}(t') \rangle \nonumber \\
&\;\;\;\;+ i\langle \hat{\rho}_{\mu,{\bf q}}(t) \rangle \langle \hat{\rho}_{\nu,{-\bf q}}(t') \rangle,
\end{align}

where $\mu,\nu=0,1$ and $\hrho_{\mu,{\bf q}} = \frac{1}{\sqrt{N}}\sum_i e^{-i{\bf q}\cdot {\bf r_i}} \hrho_{\mu,i}$ with $\rho_{\mu,i} \equiv \hat{\Psi}^\dagger_i \boldsymbol{\sigma}_\mu \hat{\Psi}_i$ and $\hat{\Psi}_i = [\hc_{i,c}\;\;\hc_{i,v}]^T$.
The retarded part of this function is the susceptibility (the response function).
At $T=0$ in the present model, the state with the valence band fully occupied ($\equiv|\Phi_0 \rangle$) is the ground state when the band gap is sufficiently large.
Therefore, 
\begin{align}
G_{cc,{\bf k}}(t,t') = 0 \; (\text{for } t\prec t'), \;\;\;\; G_{vv,{\bf k}}(t,t') = 0 \; (\text{for } t'\prec t). \label{eq:G_cond}
\end{align}
Here $t\prec t'$ indicates that $t'$ appears later than $t$ in terms of the contour ordering.
In addition, the single particle Green's function within the MF theory is exact at $T=0$ in this model, since 
$\hat{c}^\dagger_{c,\bf k}|\Phi_0 \rangle$ and $\hat{c}_{v,\bf k}|\Phi_0 \rangle$ are also eigenstates and 
the corresponding eigen energies (measured from the ground state energy) are $\epsilon_{c}({\bf k})+U+D_c$ and $-\epsilon_{v}({\bf k})-\Delta_v$.

 \begin{figure}[b]
  \centering
    \hspace{-0.cm}
    \vspace{0.0cm}
   \includegraphics[width=85mm]{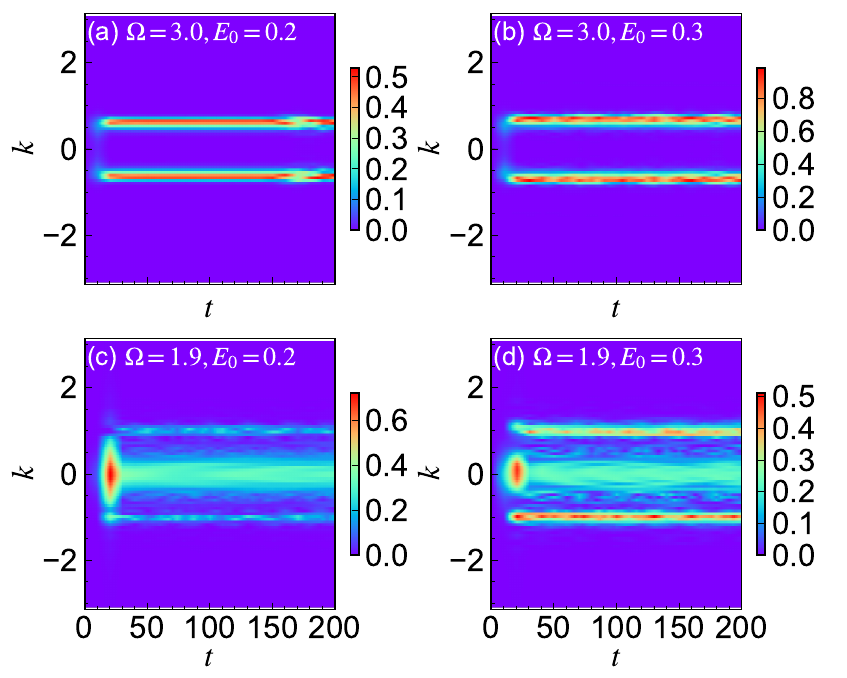} 
  \caption{Time-evolution of the momentum distribution of the conduction-band electrons ($n_c(k)$) for various indicated conditions within tdMF. 
  }
  \label{fig:nk_MF}
\end{figure}
 \begin{figure*}[t]
  \centering
    \hspace{-0.cm}
    \vspace{0.0cm}
   \includegraphics[width=170mm]{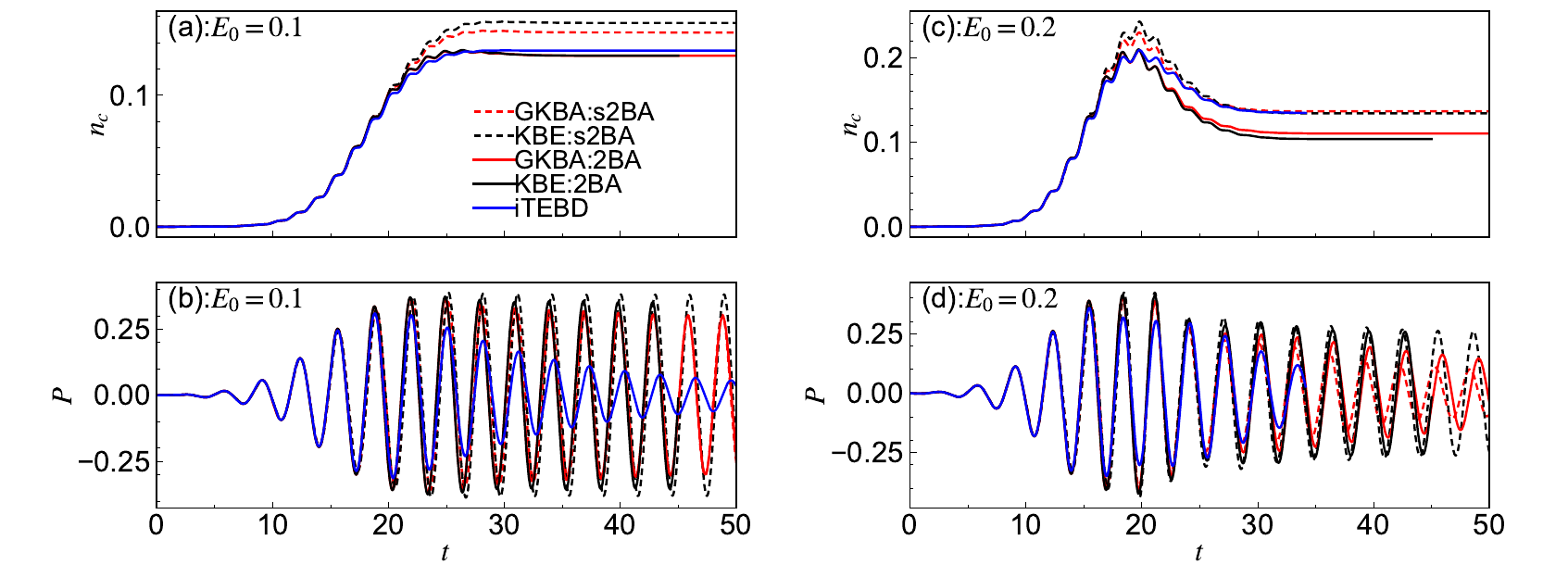} 
  \caption{Comparison of the density of conduction band electrons and the polarization among s2BA, 2BA, GKBA+s2BA, GKBA + 2BA and iTEBD for $\Omega=1.9$.
   Panels (a),(b) are for $E_0=0.1$ and (c,d) are for $E_0=0.2$.}
  \label{fig:Compare_U2Ome19E01_02_appendix}
\end{figure*}
Now we consider the diagrammatic expression for $\chi_{\mu\nu}(t,t';{\bf q}) $ in terms of the full electron Green's functions.
The expression consists of a) the ladder diagrams (Fig.~\ref{fig:diagrams_fig}(a)), 
b) diagrams which include ring diagrams of the type shown in Fig.~\ref{fig:diagrams_fig}(b), and c) the ladder-like diagrams, which include at least one crossing of the interaction lines (Fig.~\ref{fig:diagrams_fig}(c)).
However, one can show that the contributions from b) and c) are zero at $T=0$ in the present model,
because of Eq.~(\ref{eq:G_cond}).
A ring diagram consists of either $G_{cc}(t,t')$ or  $G_{vv}(t,t')$, since $G_{cv}(t,t')=G_{vc}(t,t')=0$. 
When we write the time of the vertices on the ring as $t_1(=t_{N+1}),t_2,.. t_{N}$, 
both $t_{i+1}\succ t_i $ and $t_{i+1}\prec t_i$ must appear because of the periodic boundary condition.  
Hence, according to Eq.~(\ref{eq:G_cond}), the ring contribution should always vanish.
For the ladder-like diagrams, let us write the times of the vertices in the lower lines as $t_1,t_2,,,t_N$ and those on the upper lines as $t'_1,t'_2,,,t'_N$.
The elements of  $[t_1,t_2,,,t_N]$ and those of $[t'_1,t'_2,,,t'_N]$ are identical since the interaction is instantaneous. 
To get a nonzero value for the lower part one needs $t \succ t_1 \succ t_2 \succ \ldots \succ t_N \succ t'$ or $t \prec t_1 \prec t_2 \prec,,,\prec t_N \prec t'$,
while a nonzero upper part requires $t \succ t'_1 \succ t'_2 \succ,,,\succ t'_N \succ t'$ or $t \prec t'_1 \prec t'_2 \prec \ldots \prec t'_N \prec t'$.
In a ladder with crossed interaction lines these two conditions cannot be simultaneously satisfied, so that the contributions from diagrams of the type shown in Fig.~\ref{fig:diagrams_fig}(c) also vanish.
Therefore, only ladder diagrams can give a nonzero contribution to $\chi_{\mu\nu}$.

One can show that the summation of all the ladder diagrams leads to Eq.~(\ref{eq:rpa_like}) with Eq.~\eqref{eq:chi0_noneq}, where 
$\mathcal{G}_k$ is the exact equilibrium Green's function at $T=0$.
Hence, the susceptibility evaluated from the MF dynamics at $T=0$ is exact.

Alternatively, one can use the expression of the response function following Eq.~\eqref{eq:chi_contour}\cite{stefanucci_nonequilibrium_2013,Murakami2016PRBa,Murakami2016PRBb} to show that the response function obtained by the tdMF, s2BA and 2BA is exact.
The ladder diagrams originate from the functional derivative of the Fock diagram ($\frac{\delta_{\mathcal C} \Sigma_F[G]}{\delta_{\mathcal C} F_{\rm ex}}$).
Since the Fock term Eq.~\ref{eq:Fock} is included in all of these methods, the corresponding susceptibility also includes the ladder diagrams.
On the other hand, the equilibrium Green's functions described by these approximations are exact.
(From Eq.~\eqref{eq:G_cond}, the correlated part of the self-energy should be zero at $T=0$.)
Hence the diagrams other than the ladder diagrams in the susceptibility vanish for the same reason as discussed above.
Therefore,  tdMF, s2BA and 2BA also produce the exact response functions at $T=0$ in this model.

 As for the GKBA, $\rho_{cc,{\bf k}}=0,\rho_{vv,{\bf k}}=1, \rho_{cv,{\bf k}}=0,\rho_{cv,{\bf k}}=0$  yields $\bSig^{</>}_{\rm corr}[\tilde{\bG}]=0$.
 Hence this is a steady state solution and the adiabatic switching of the interaction leads to this state.
 When the excitation with the off-diagonal field ($\sum_{i,a} \hc^\dagger_{i,a} \hc_{i,\bar{a}}$) is applied to this ground state, the linear response in $\hc^\dagger_{i,b} \hc_{i,b}$ should be zero, since the corresponding linear response function ($\simeq\langle \hc^\dagger_{i,b}(t) \hc_{i,b}(t) \hc^\dagger_{i,a}(0) \hc_{i,\bar{a}}(0)\rangle$) is zero due to the 
 conservation of particles in each orbital. 
 Hence the response of $\rho_{k,cc}$ and $\rho_{k,vv}$ against the field starts from $\mathcal{O}(E_0^2)$.
 Therefore, $\tilde{G}_{cc}^<,\tilde{G}_{vv}^> =  \mathcal{O}(E_0^2)$ and  $\tilde{G}_{cv},\tilde{G}_{vc}  = \mathcal{O}(E_0)$. 
 Using these facts and directly evaluating $\bSig^{</>}_{\rm corr}[\tilde{\bG}]$, one can show that all components in $\bSig_{\rm corr}$ behave as $\mathcal{O}(E_0^2)$.
 Hence in the linear response regime, the collision integral is still zero and the time evolution is the same as in the MF theory.

\section{Momentum distribution from tdMF}
\label{appendixB}
In Fig.~\ref{fig:nk_MF}, we show the momentum distribution of the conduction-band electrons ($n_c(k)$) evaluated with tdMF.
For the above-gap excitation ($\Omega=3.0$), the charges are excited at finite momentum and are stuck there after the excitation because of the absence of scattering in tdMF.
For the resonant excitation, there emerges some finite occupation around $k=0$ and $k=\pm1.0$.
The occupation around $k=0$ corresponds to the direct excitation, while that around $k=\pm1.0$ corresponds to absorption of two photons.

\section{Effects of the exchange term}\label{appendixC}

In Fig.~\ref{fig:Compare_U2Ome19E01_02_appendix}, we compare the time-evolutions described by s2BA, 2BA, GKBA+s2BA, GKBA + 2B and iTEBD for $\Omega=1.9$
to see the effect of the exchange term (Eq.~\eqref{eq:sigma2b_x}).
While there are rather clear effects on the number of photo-carriers, the inclusion of the exchange term does not generally lead to a quantitative improvement of the results.
As for the effect on the time evolution of the polarization, it seems less prominent and again there is no clear improvement associated with the exchange term.

\bibliographystyle{apsrev4-1}
\bibliography{ultrafastexciton}

\end{document}